\documentclass[%
 twocolumn,
 amsmath,amssymb,
aps,
]{revtex4-2}

\usepackage{graphicx}
\usepackage{dcolumn}
\usepackage{bm}
\usepackage{subcaption}
\usepackage{xcolor}
\usepackage{hyperref}
\hypersetup{
    colorlinks=true,
    linkcolor=blue,
    filecolor=magenta,      
    urlcolor=cyan,
    pdftitle={Overleaf Example},
    pdfpagemode=FullScreen,
    }

\newcommand{\RR}{\mathbb{R}}
\newcommand{\CC}{\mathbb{C}}

\DeclareMathOperator*{\argmax}{\arg\!\max}
\DeclareMathOperator*{\argmin}{\arg\!\min}

\begin{document}

\preprint{APS/123-QED}



\title{Feel the Force: From  Local Surface Pressure Measurement to Flow Reconstruction in Fluid-Structure Interaction} 


\author{Colin Rodwell}
\author{Kumar Sourav}
\author{Phanindra Tallapragada}%
 \email{ptallap@clemson.edu}
\affiliation{%
 Department of Mechanical Engineering, Clemson University, Clemson 29634, SC, USA\\
}%




\date{\today}

\begin{abstract}

Drawing inspiration from the lateral lines of fish, the inference of flow characteristics via surface-based data has drawn considerable attention. The current approaches often rely on analytical methods tailored exclusively for potential flows or utilize black-box machine learning algorithms to estimate a specific set of flow parameters. In contrast to a black box machine learning approach, we demonstrate that it is possible to identify certain modes of fluid flow and then reconstruct the entire flow field from these modes. We use Dynamic Mode Decomposition (DMD) to parametrize complex, dynamic features across the entire flow field. We then leverage deep neural networks to infer the DMD modes of the pressure and velocity fields within a large, unsteady flow domain, employing solely a time series of pressure measurements collected on the surface of an immersed obstacle. Our methodology is successfully demonstrated to diverse fluid-structure interaction scenarios, including cases with both free oscillations in the wake of a cylinder and forced oscillations of tandem cylinders, demonstrating its versatility and robustness.
\end{abstract}

\maketitle


\section{Introduction}

A body or structure immersed in a fluid flow is subject to hydrodynamic forces due to fluid-structure interaction. Different flow patterns in general lead to different pressure distributions on the surface of the body, and it is natural to ask if the fluid flow can be inferred solely based on pressure or velocity measurements on the surface of the immersed body. Many marine animals seemingly have an ability to at least sense and localize disturbances, if not generate detailed understanding flow patterns based on non-visual information such as pressure measurements.  A well known example of such ability is the schooling of fish \cite{li2020vortex, cooke2022movement} which requires real-time understanding of the positions and directions of neighboring fish, and can be performed by blind fish using only the ``lateral line'', a sophisticated line of biological pressure and velocity sensors located along the sides of many fish \cite{liu2016review, pitcher1976blind}. Inspired by this, much research has been focused on developing ``artificial lateral lines,'' which mimic biological lateral lines using artificial sensors. Some efforts using artificial lateral lines have succeeded in determining specific parameters of the flow, such as the location and movements of sources and dipoles, using analytical approaches by assuming potential flow \cite{yen2020localization,abdulsadda2013underwater} and by employing learning-based methods \cite{wolf2020three, rt_acc_2022}. In other related work; black-box neural networks have  been used to classify and predict wake features using  the fluid velocity field around oscillating foils in \cite{kanso_bb_2018, franck_aiaa_2021} and variational autoencoders have been used to reconstruct flows using velocity measurements in the flow domain \cite{dubois_jcp_2022}. Shallow neural networks have also been used to reconstruct flows from sensor measurements on the surface of a body for scenarios such as flow past a cylinder such as in \cite{brunton_prsa_2020}. Neural networks have also been used to predict or classify wake features such as the Strouhal number of a wake based solely on the kinematics of a trailing body immersed in the wake in \cite{pt_bb_2021, rpt_bb_2023}.   Physics-informed neural networks, have also been used to solve the inverse problem of finding the pressure distribution on a body given sparse measurements of the velocity field of the fluid surrounding the body in \cite{raissi2019deep}. However, designing a more general framework for understanding the ambient flow dynamics based solely on surface measurements remains an open problem largely due to the high dimension and unsteady nature of the fluid flow.

The question addressed in this paper then is, ``can the flow field around the body be reconstructed knowing only pressure measurements at a few points on a body immersed in the fluid?" We show that this can be done by a combination of dynamical systems tools and machine learning. Instead of directly reconstructing a flow field using black-box machine learning, we first show that the modal decomposition of a sparsely sampled pressure field on the surface of the body can be correlated to the modes of the fluid flow field via supervised learning by shallow neural networks. The full flow field can then be reconstructed using the identified modes. The technique used for the modal decomposition is Dynamic Mode Decomposition (DMD). We demonstrate this using two fluid-structure interaction examples, where pressure measurements on a trailing body in the wake of leading body are used to reconstruct the flow in a domain with a length scale that is many times bigger than the body length. 

Dynamic Mode Decomposition (DMD) \cite{schmid2010dynamic,schmid2011applications} offers an approach to approximate an unsteady flow by modeling it as a superposition of linear modes. Because many of these modes are often insignificant to the dynamics, only a few modes can typically reconstruct the evolution of the flow with high fidelity, allowing a reduction in the temporal dimension. These modes often have physical meaning, which makes DMD a useful tool for extracting and elucidating dominant flow structures and associated dynamics \cite{schmid2010dynamic,schmid2011applications,rowley2009spectral}. This low dimensionality and practical usefulness raise the possibility that estimating the DMD modes of the surrounding fluid based on surface pressure measurements may be both tractable and useful.

A procedure for determining the DMD modes around a body based on surface pressures was first presented in \cite{bright2013compressive}, where the fluid modes at an unknown Reynolds number were selected from a dictionary of known modes (calculated for a range of Reynolds numbers) based on which modes could most accurately explain the surface measurements. A different approach was considered in \cite{paley_jgcd_2019}, where the modes of the flow were known, and the correct superposition of those modes to explain the flow at a given time was selected from surface pressure measurements using a filtering approach. In this work, we instead take a parametric approach that requires no dictionary of modes to operate. This is done by training a neural network that estimates the dominant DMD modes of the flow pressure and velocity, using the DMD modes calculated only using pressure measurements on the surface as input. The idea of using a neural network with DMD modes to reconstruct the fluid field has been explored very recently in \cite{eivazi2020deep,zhang2023nonlinear}; however, these works use autoencoders to map the velocity of the entire fluid field data to a latent space, perform DMD in the latent space, and map the results back to the same velocity field. By contrast, in this paper DMD is performed on the surface pressure or the field velocity and pressure directly, and the mapping is from the surface DMD modes to the DMD modes of the fluid velocity field. 

While the primary motivation for the problem investigated is related to sensing by fish-like underwater robots, other engineering applications are possible. Fluid-structure interaction such as in vortex induced oscillations \cite{williamson2004vortex}, wake-induced vibrations (WIV) and forced oscillations in tandem cylinder arrangements, see for example \cite{zdravkovich1985flow, mahir1996vortex, meneghini2001numerical, mittal2001flow} is of critical relevance to structural integrity in aircraft design, ship design, submersible vehicles, offshore structures and heat exchangers, see for example \cite{jing2022wake,chen2023effect,awadallah2023numerical,assi2010suppression, li2020investigation}.  The increasing ubiquity of sensors and computing opens up possibilities for near real time sensing, estimation and control of local flow and structural response. This will require a framework of estimating flow field from onboard structures immersed in the flow.  

The remainder of this paper is structured as follows: After the Introduction, Section II delves into the problem setup for wake-induced vibration (WIV) and forced oscillations, elaborating on the governing equations for both fluid dynamics and structural motion. This section also outlines the computational mesh details and studies on mesh independence and validation. Section III reviews the Dynamic Mode Decomposition (DMD) and discusses its relevance in the context of the flow reconstruction. Section IV describes the proposed method employed for flow reconstruction. Section V presents the results concerning the application of DMD and flow reconstruction on both WIV and forced oscillation systems. Finally, Section VI summarizes the key findings of this research in the Conclusions.

\section{Numerical Simulation of Fluid-Structure Interaction}

We consider two examples of fluid-structure interaction. In the first example  a circular cylinder mounted on a spring-damper system is free to oscillate laterally in the wake of a stationary square prism. In the second example, two circular cylinders are forced to oscillate out of phase with varying amplitudes, aiming to increase flow complexity. A  description of the numerical simulation of the two problems  follows.

\subsection{Fluid-Structure interaction simulations setup}

\begin{figure} [ht]
\begin{center}
\includegraphics [width=0.98 \hsize]{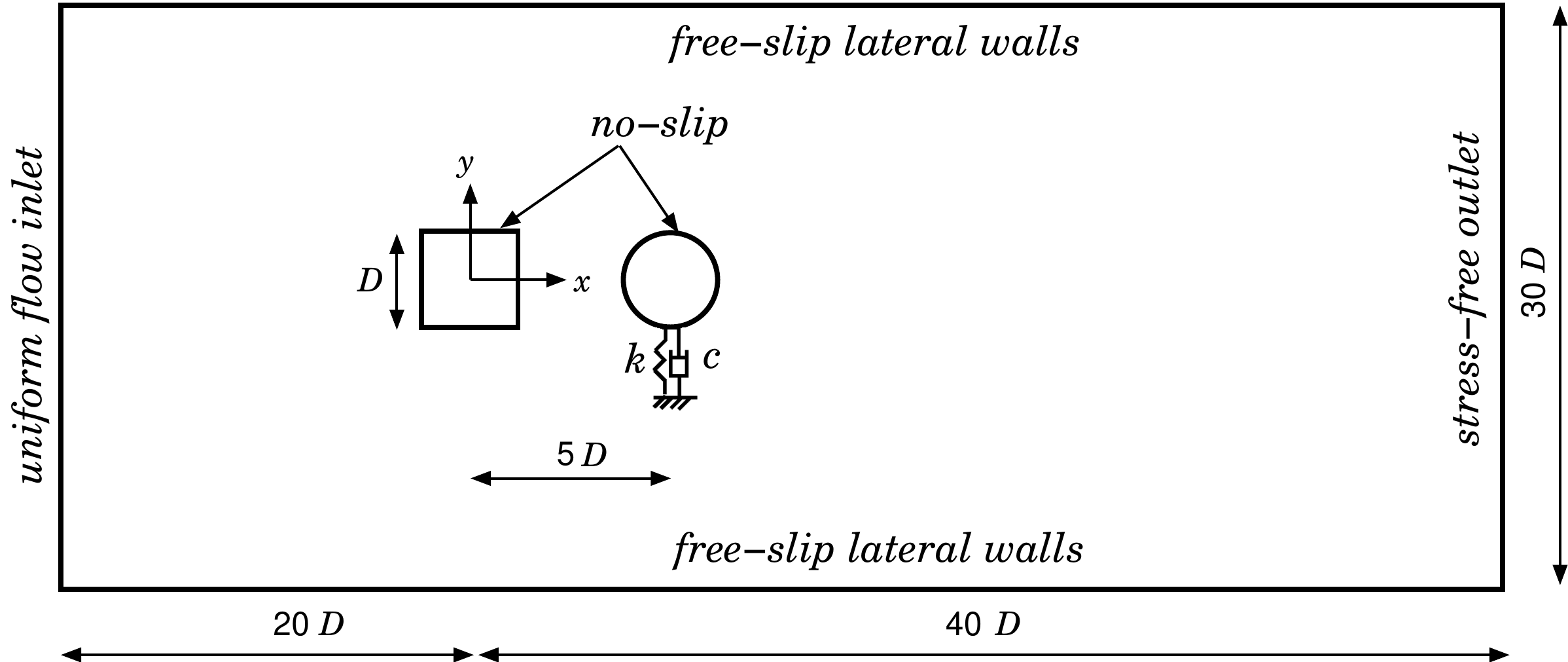}
\end{center}
\vspace{-.2 in}
\caption{Schematic representation of the system setup for the study of wake-induced transverse oscillations of a circular cylinder, positioned 5$D$ downstream of a stationary square cylinder. Both cylinders have a cross-stream length, `$D$.' The circular cylinder is attached to a linear spring and damper system.}
\label{problem_setup}
\end{figure}

Figure~\ref{problem_setup} illustrates the computational domain setup for the first problem with a downstream circular cylinder placed in the wake created by an upstream square prism. While the circular cylinder is restricted to cross-flow oscillations, the square prism remains stationary. Both bodies have the same characteristic length, `$D$,' which equates to the prism's side length and the circular cylinder's diameter. Their center-to-center gap is denoted by the gap ratio ($S/D$) and is fixed at a value of 5, which is above the critical value suggested in a number of studies \cite{zdravkovich1977interference, papaioannou2008effect, kumar2019steady, yang2019critical}. All simulations take place within a two-dimensional space and maintain a constant Reynolds number ($Re$) of 100. The downstream cylinder operates as a one-degree-of-freedom (1-DOF) mass-damper-spring system with a specified mass ratio ($m^*$) of 10.0 and a damping ratio ($\zeta$) of 0.2. By keeping both $Re$ and cylinder diameter consistent, the reduced velocity, $U^* = \frac{U}{{f_n}D}$ (with $f_n$ being the natural frequency of the oscillating circular cylinder) is varied across simulations from 1 to 15  by modulating the oscillator's natural frequency.

The computational fluid domain, shown in Figure \ref{problem_setup}, spans a $60D\times30D$ rectangular area, utilizing a rectangular coordinate system with the origin set at the center of the upstream body. This domain is flanked symmetrically on the top and bottom by boundaries spaced 30$D$ apart, resulting in a blockage ratio ($B$) of 3.33$\%$. The distance between the lateral boundaries has little impact on the flow field around the cylinders if the blockage ratio is less than 5$\%$, as established by \cite{sourav2019transition, zhu2019wake,zhu2019flow, zhang2022fluid,sourav2022simultaneous,kumar2023vortex, sen2011free, sourav2020vortex}. Regardless of changes in cylinder positions or flow attributes, the fluid domain's upstream and downstream limits remain constant in all simulations at 20$D$ and 40$D$ from the origin, respectively. The cylinders' surfaces observe a no-slip condition, ensuring no relative motion between the fluid and the cylinder. The free-stream velocity at the upstream boundary is characterized by $u = U$ and $v = 0$. The downstream boundary enforces a zero gradient for flow velocities, facilitating the smooth exit of the fluid. The top and bottom boundaries employ a slip wall condition, defined by $\frac{\partial u}{\partial y} = 0$ and $v = 0$, mimicking a shear-free environment and mitigating interference with the flow dynamics. 

\begin{figure} [ht]
\begin{center}
\includegraphics [width=0.96 \hsize]{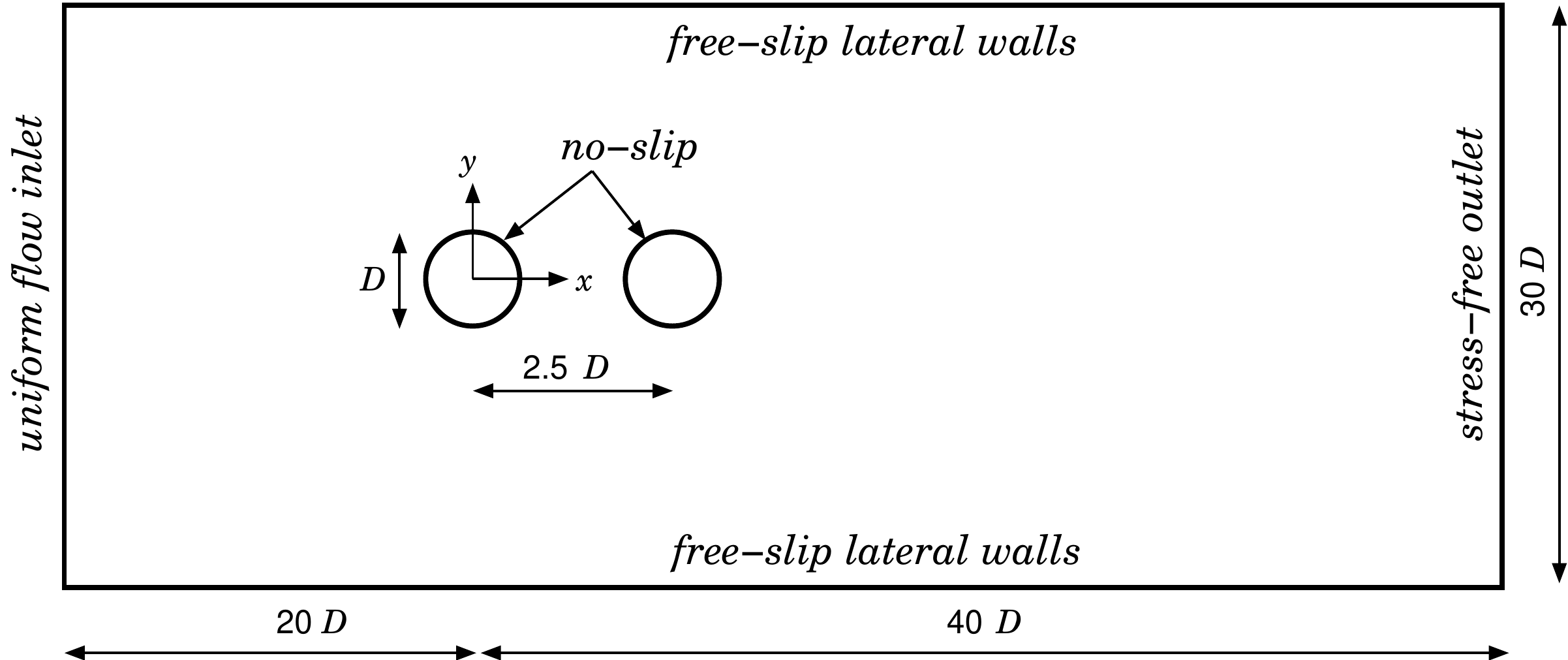}
\end{center}
\vspace{-.2 in}
\caption{Schematic representation of the system setup for the study of forced transverse anti-phase oscillations of tandem circular cylinders, separated by a distance of 2.5$D$. Both cylinders have a cross-stream length `$D$'. The origin of the Cartesian coordinate system aligns with the geometric center of the upstream circular cylinder.}
\label{problem_setup_forced}
\end{figure}

Figure \ref{problem_setup_forced} is a schematic of the  computational domain tailored for the study of forced transverse oscillations of two tandem circular cylinders. Located at the geometric center of the upstream circular cylinder, the Cartesian coordinate system's origin serves as a reference. The cylinders, separated by a distance of 2.5$D$ (yielding $S/D = 2.5$), undergo anti-phase oscillations, a setup crafted to accentuate the intricacies of the flow. The oscillation amplitude $A$ is systematically adjusted from 0.1 to 1.0 in increments of 0.1 while maintaining a consistent oscillation frequency for both cylinders at 1.04 rad/s. As the dimensions and boundary conditions of this computational domain mirror those of the free oscillations problem, further elaboration is omitted for conciseness. \textit{For the sake of conciseness, discussions related to independence studies, mesh details, and validation will be presented exclusively for the first problem henceforth.}

\subsection{Governing Equations and Solution Methodology}

The computational analyses leveraged two-dimensional direct numerical simulations performed using the open-source computational fluid dynamics (CFD) platform, OpenFOAM, accessible at \url{www.openfoam.org}. OpenFOAM utilizes the finite volume method for discretizing continuum mechanics problems, including the unsteady Navier-Stokes equations \eqref{momentum_eqn} and \eqref{continuity_eqn}. These were discretized in conjunction with the Pressure Implicit with Splitting of Operators (PISO) algorithm. A fourth-order cubic interpolation scheme ensured high accuracy for spatial derivatives by discretizing the convective term in the equations. On the other hand, the diffusion term was discretized using a second-order linear scheme. The derivative term's temporal discretization was achieved through a blended scheme combining the second-order Crank-Nicolson scheme and the first-order Euler implicit scheme, providing a high degree of accuracy in temporal resolution while ensuring numerical stability.
\begin{eqnarray}
\label{momentum_eqn}
    \rho \left( \frac{\partial \bf{u}}{\partial t} + \bf{u} \cdot  \bf{\nabla} \bf{u} \right) = -\bf{\nabla} p + \mu \bf{\nabla}^2 \bf{u}, \\
\label{continuity_eqn}
    \nabla \cdot \bf{u} = 0.
\end{eqnarray}
The hydrodynamic forces acting on the cylinder surface are derived directly from solving the Navier-Stokes equations, subsequently triggering the vibrational response of the circular cylinder. The governing equation for the cylinder's cross-flow oscillations can be expressed as
\begin{equation}
\label{cyl_motion}
m\Ddot{y}+c\Dot{y}+ky = f_L.
\end{equation}
Here, $m$ represents the mass of the circular cylinder, and $\Ddot{y}$, $\Dot{y}$, and $y$ respectively denote the cylinder's transverse acceleration, velocity, and displacement. The system damping is designated by $c$; the spring stiffness is represented by $k$, and $f_L$ corresponds to the unsteady lift force. This equation integrates the fluid dynamics with the structural dynamics, thereby providing a comprehensive model for the study of the cylinder's oscillatory behavior within the fluid flow.

The simulations were carried out in an iterative manner, alternating between solving for the fluid field and the structural response at each time step. Initially, the velocity and pressure distributions in the fluid domain were determined, followed by the computation of the drag and lift forces through the integration of the pressure and shear stress on the cylinder surface. The calculated hydrodynamic force was then substituted into Equation \ref{cyl_motion}, leading to the computation of the cylinder's transverse displacement ($y$) using an enhanced fourth-order Runge-Kutta method. This displacement information subsequently dictated updating the computational grids, thereby establishing a new mesh for the fluid field calculation in the following time step. This iterative process continued until the system's dynamic behavior reached stabilization and a sufficient number of cyclical results were accumulated. The selected time steps, denoted by $\Delta t$, for each case adhered to the Courant-Friedrichs-Lewy (CFL) condition, maintaining a number below 0.85 across the entire computational domain to ensure numerical stability and the accuracy of simulation results. Note that for the free oscillations scenario, both the fluid-flow equations (Equations 1 and 2) and the structural motion equation (Equation 3) are tackled. In contrast, the forced oscillation problem exclusively focuses on resolving the fluid-flow equations because the displacements of both cylinders are prescribed.

\begin{figure} [ht]
\begin{center}
\includegraphics[width=0.96 \hsize]{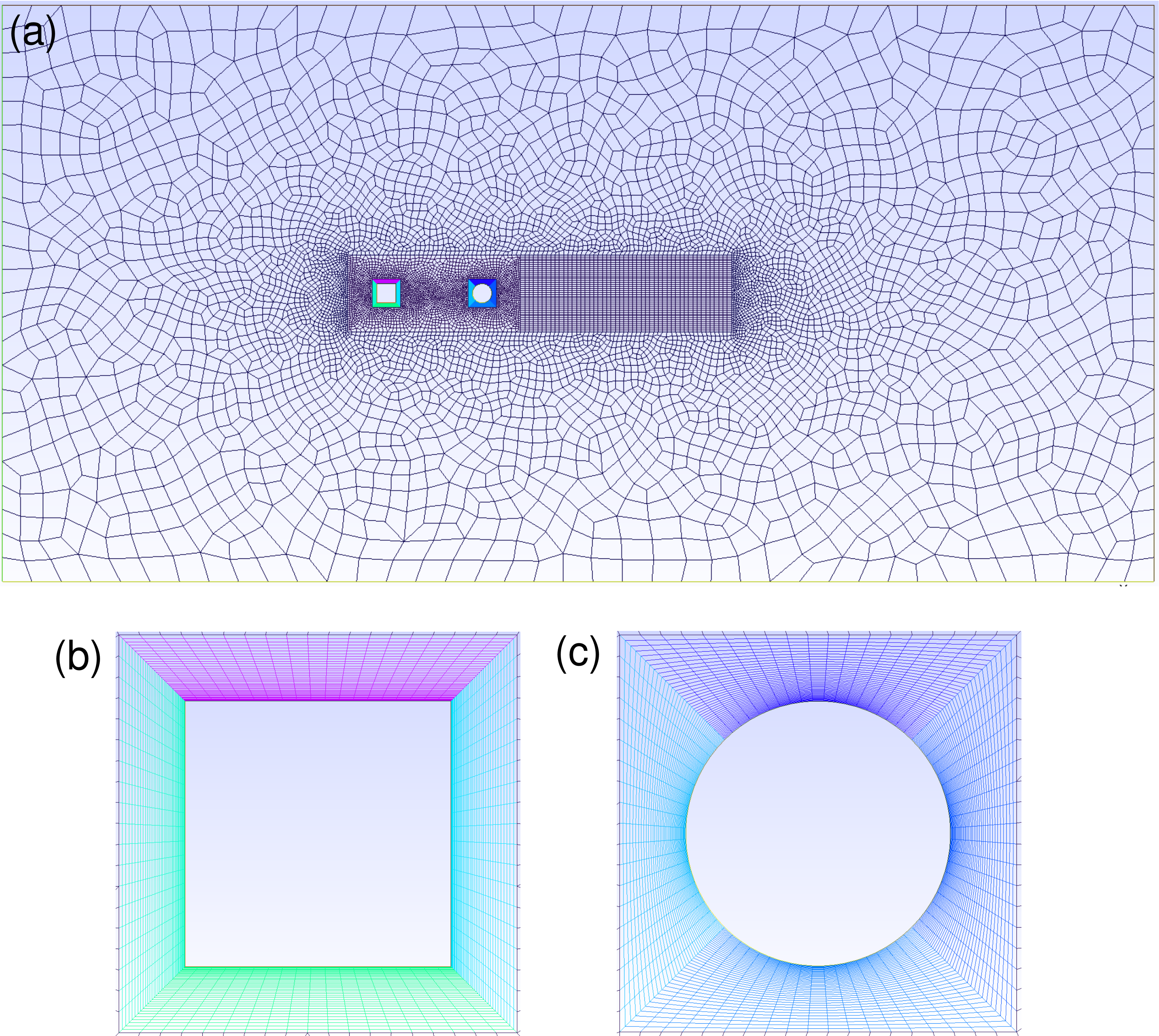}
\end{center}
\vspace{-.2 in}
\caption{(a) The unstructured finite volume mesh in the computational fluid domain of size 60$D$ $\times$ 30$D$. The mesh near the cylinders and the wake region is structured. (b, c) Zoomed-in view of the structured mesh near the square and circular cylinders.}
\label{mesh}
\end{figure}

\subsection{Mesh Details}

As depicted in Figure \ref{problem_setup}, the rectangular computational fluid domain is discretized using an unstructured finite volume mesh composed of 32,434 nodes. Spatial variation in mesh density is employed, with higher density in regions proximal to the cylinders and coarser density towards the domain boundaries, as shown in Figure \ref{mesh}(a). Detailed views of this meshing near the upstream and downstream cylinders are shown in Figures \ref{mesh}(b) and \ref{mesh}(c), respectively, where 80 grid points define each cylinder. The cylinders are encapsulated within their respective square blocks of dimensions $1.5D \times 1.5D$ to minimize projection error during oscillations. Structured, non-uniform meshes discretize the regions between each cylinder and its enclosing square block. The initial grid line from the cylinder surface is set at a distance of 0.005$D$ and extends geometrically towards the block boundaries with a progression ratio of 1.05, leading to a line segment between the cylinder and the block consisting of 44 grid points. During oscillations, the square block moves concurrently with the cylinder, preserving the internal mesh structure, while the surrounding mesh deforms in response to the transverse motion of the cylinder. To capture the wake dynamics effectively, an additional finely meshed rectangular region, characterized by dimensions $13D \times 4D$, is established in the wake of the downstream cylinder.

\subsection{Mesh Convergence and Validation}

Mesh convergence tests are crucial in computational studies to ensure that the results are sufficiently independent of the mesh resolution. In this study, three different mesh densities, namely M1 (fine), M2 (medium), and M3 (coarse), were utilized. These meshes were evaluated at flow conditions defined by a Reynolds number ($Re$) of 100 and a reduced velocity ($U^*$) of 9.


Table~\ref{table_mesh} presents a comparative analysis of characteristic flow and vibration metrics across the three mesh densities. A close evaluation reveals distinct differences between the results from the fine mesh, M1, and those from the medium (M2) and coarse (M3) meshes. Notably, the outcomes from M2 and M3 align closely, demonstrating high consistency. This level of agreement, coupled with the limited deviation between these meshes, underscores the capability of the medium mesh, M2, to strike an optimal balance between computational efficiency and accuracy. As a result, M2 was selected as the preferred mesh for all subsequent computations.
\begin{table}[ht]
\begin{center}
\begin{tabular}{c c c c c c c} \hline
    Mesh & Nodes & $Y_{max}/D$ & $C_{l_{rms}}$ & $C_{d_{avg}}$ & $F_y$ & $F_{C_{l}}$\\ \hline
    M1  & 56,814 & 0.5988 & 0.6637 & 0.7037 & 0.1136 & 0.1136  \\ 
    \textbf{M2}  & \textbf{32,434} & \textbf{0.5936} & \textbf{0.6752} & \textbf{0.7122} & \textbf{0.1152} & \textbf{0.1152}   \\ 
    M3  & 20,657 & 0.5346 & 0.7303 & 0.7773 & 0.1237 & 0.1237   \\ \hline
\end{tabular}
\caption{Comparison of characteristic flow and vibration quantities for three different mesh densities. The data presented is for a vibrating circular cylinder (with a mass ratio $m^*$ = 10) that is free to oscillate in the wake of a stationary square cylinder exposed to uniform flow conditions at $Re$ = 100 and $U^*$ = 9.}
\label{table_mesh}
\end{center}
\end{table}

\begin{figure} [ht]
\centering
\captionsetup[subfigure]{labelformat=empty}
\begin{subfigure}[b]{0.99\hsize}
    \centering
    \includegraphics [scale=0.48, angle = 0]{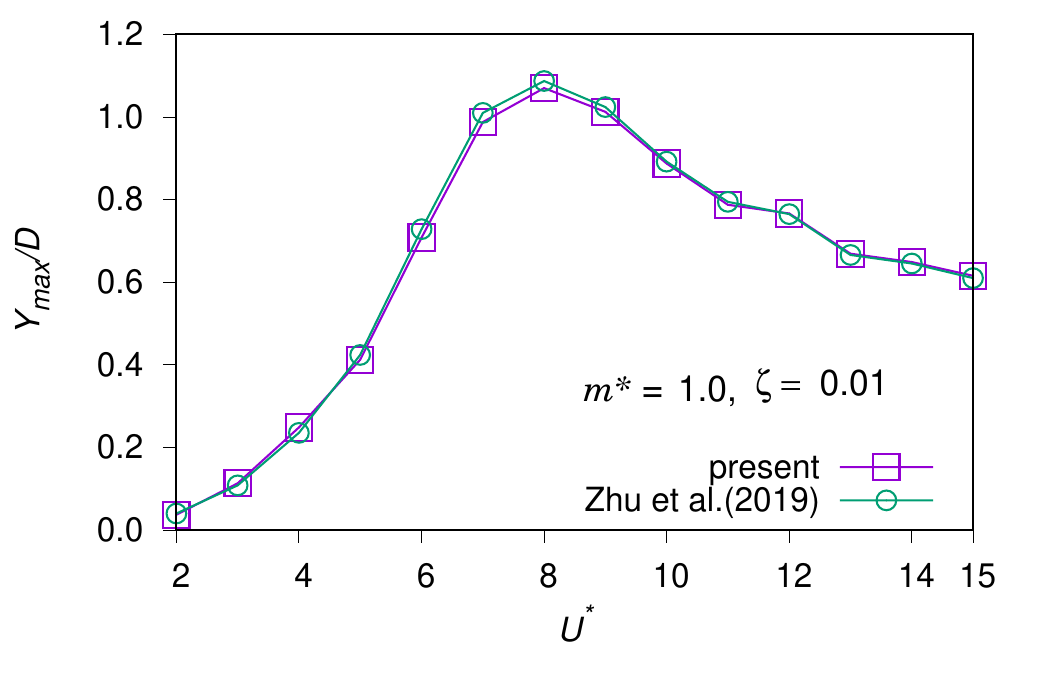}
    \caption{$(a)$}
\end{subfigure}
\begin{subfigure}[b]{0.99\hsize}
    \centering
    \includegraphics [width=0.98\hsize]{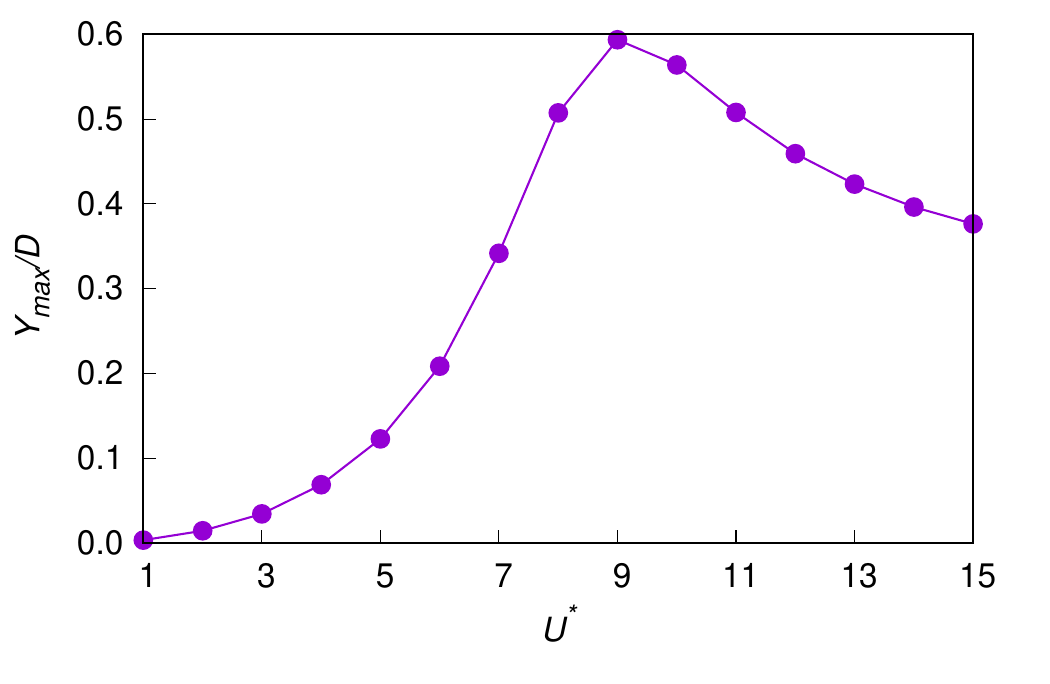}
    \caption{$(b)$}
\end{subfigure}

\caption{(a) Validation of the numerical model for wake-induced vibrations of a circular cylinder situated in the wake of a stationary square cylinder separated by a distance of 5$D$. The normalized maximum oscillation amplitude ($Y_{max}/D$) of the downstream circular cylinder ($m^* = 1$) is compared with that reported by \cite{zhu2019wake}. (b) Wake-induced vibrations of a circular cylinder of $m^* = 10$ at $Re$ = 100 and $\zeta = 0.2$: maximum transverse oscillation amplitude ($Y_{max}$) normalized with $D$ at $S/D$ = 5.}
\label{validation}
\end{figure}

The transverse oscillation response of a circular cylinder, normalized as \(Y_{max}/D\), oscillating in the wake of a stationary square cylinder is depicted in Figure~\ref{validation}(a). This response is juxtaposed with the findings presented by \cite{zhu2019wake}. The simulations were executed at a Reynolds number (\(Re\)) of 100, with the reduced velocity (\(U^*\)) spanning a range from 2 to 15. The vibrating downstream circular cylinder maintains a mass ratio (\(m^*\)) of 1 and a damping ratio (\(\zeta\)) of 0.01 for the system. A close examination reveals a commendable concordance between the computed data and the results from \cite{zhu2019wake}. Therefore, our projected response data for wake-induced vibrations of a circular cylinder aligns satisfactorily with the findings of \cite{zhu2019wake}.

\subsection{Simulation Results}

\begin{figure}
    \centering   
    \captionsetup[subfigure]{labelformat=empty}
    \begin{subfigure}[b]{0.99\hsize}
        \centering
        \includegraphics[width=.98\hsize]{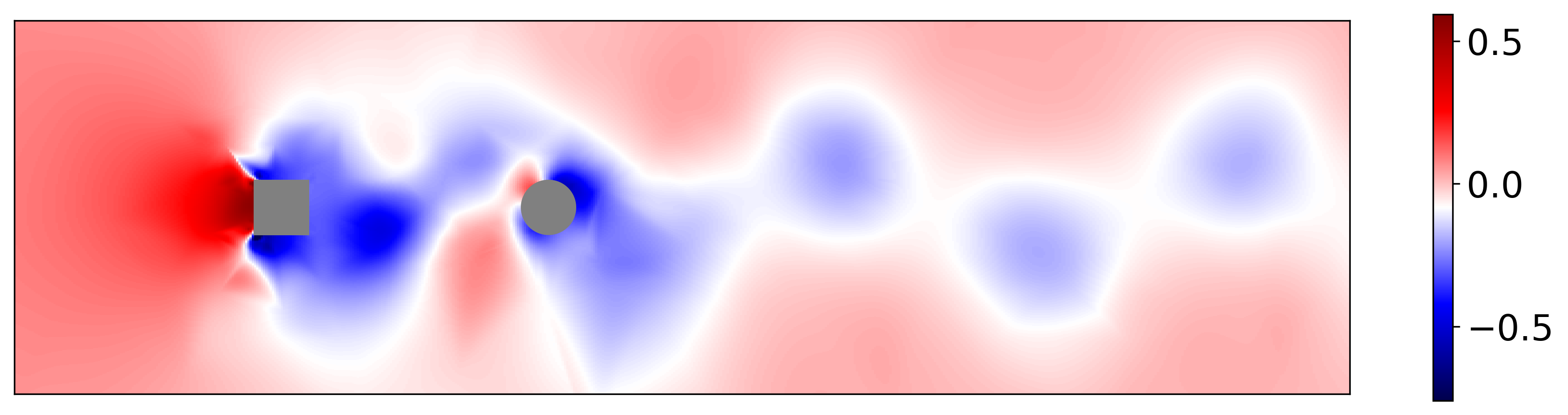}
        \caption{$(a)$}
    \end{subfigure}
    \begin{subfigure}[b]{0.49\textwidth}
        \centering
        \includegraphics[width=.98\hsize]{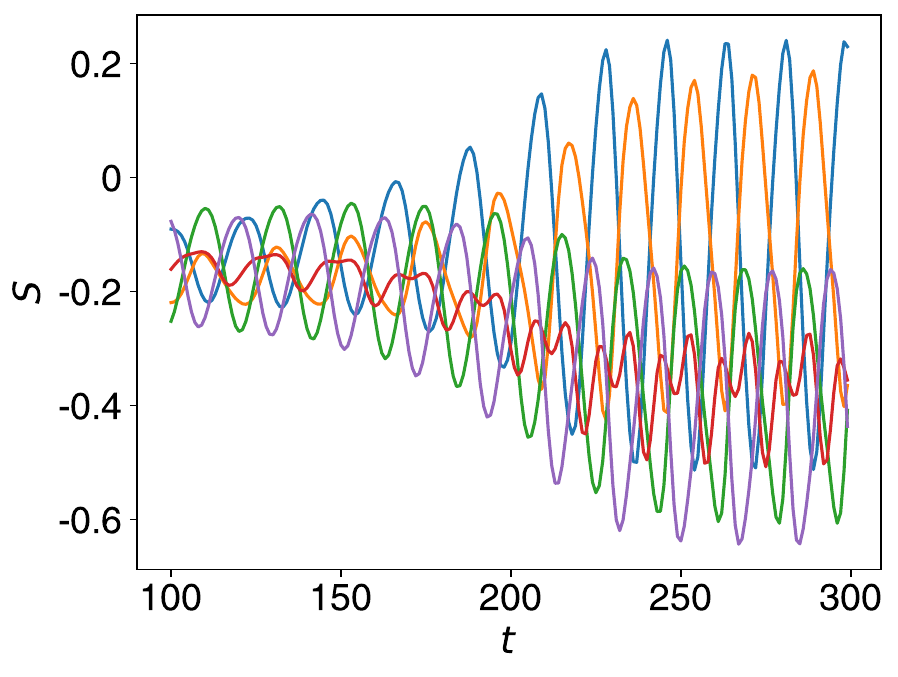}
        \caption{$(b)$}
    \end{subfigure}
    \caption{(a) The pressure field for $U^*=8$ at time $t=300$, which has settled into periodic vortex shedding. (b) The pressure at 5 evenly spaced points around the cylinder for $U^*=8$.}
    \label{fig:pressure_profile}
\end{figure}

Figure \ref{fig:pressure_profile}(a) shows the pressure field derived from the free oscillation simulation. A low-pressure oscillating wake is located behind the square body, which is then advected past the circular cylinder. This results in periodic forcing on the surface of the cylinder, visualized in Figure \ref{fig:pressure_profile}(b). Once advected past the cylinder, the wake assembles into an organized vortex street.

\section{Dynamic Mode Decomposition}
\subsection{Koopman Operator}
Consider a two-dimensional fluid flow containing one or more rigid bodies. On one body $B$, there is a distribution of pressure sensors. While the Navier-Stokes equation governs the fluid flow over a continuum of space and time, we will assume that the flow is spatially and temporally discretized, i.e. the flow domain is discretized by $N$ points $(x_i, y_i) \in \RR^2$ and $M$ evenly spaced time snapshots at times $t_n \in \RR$, and the vector-valued states are $u,v,p \in \RR^{N}$, corresponding the horizontal velocity, vertical velocity, and pressure, respectively. In the Eulerian approach considered here, each element of these vectors corresponds to a specific point $(x_i,y_i)$, which does not vary with time, though the states themselves do evolve with time.
Though the evolution of these variables is governed by a partial differential equation, there exists an unknown flow map $\mathcal{F}$ that can predict the evolution of the states after a fixed length of time $\Delta t=t_{n+1}-t_{n}$, e.g.
\begin{equation}
    (u_{n+1},v_{n+1},p_{n+1})=\mathcal{F}(u_n,v_n,p_n),
\end{equation}
where $u_n$ denotes $u(t_n)$.
It follows that by defining a generalized coordinate $z=(u,v,p) \in \RR^{3 N}$, the map can be rewritten as
\begin{equation}
    z_{n+1}=\mathcal{F}(z_n).
\end{equation}
It has been shown \cite{koopman_1931, lasota_1994, mezic_chaos_2012} that the vector-valued Hilbert space of all functions in $L^{2}$ (often called observations) of $z$, a typically infinite-dimensional space denoted $g(z)$, can also be mapped forward by an infinite-dimensional operator $\mathcal{K}$ as
\begin{equation}
    g(z_{n+1})=\mathcal{K} g(z_n).
\end{equation}
The operator $\mathcal{K}$ is linear and known as the Koopman operator, named after its inventor who derived it for conservative Hamiltonian systems in \cite{koopman_1931}. This linearity has the potential to enable the application of many linear system analysis techniques (such as modal analysis) to complex and high-dimensional nonlinear systems such as fluids and this has led to an interest in numerical approximations of the Koopman operator, see for instance \cite{rowley_jfm2009, schmid_jfm2010, mezic_chaos_2012, klus2015numerical,williams2015edmd, brunton2021modern}. In particular the Koopman operator methods have become popular following the  development of Dynamic Mode Decomposition (DMD), an algorithm that uses a subset of the observations $h(z) \subseteq g(z)$ where $h(z) \in \RR^k$ for $k<\infty$ and identifies a linear operator $K \in \RR^{k \times k}$ that satisfies
\begin{equation}\label{eqn:minimization}
K=\argmin_A \sum_{n=1}^{M-1} \|h(z_{n+1})-A h(z_n)\|_2^2 ,
\end{equation}
where $\| \cdot \|_2$ denotes the $L^2$ norm. This equation intuitively is minimizing the squared error in the prediction of the operator $A$. This finite-dimensional approximation of $\mathcal{K}$ can enable practical linear analysis while introducing some (often small) amount of error dependent on how effective $h(z)$ is as a basis. One particularly useful linear analysis technique is the computation of the eigenvalues and eigenvectors of $K$, known as the DMD modes, which elucidate key structures of the flow. In fluid dynamics problems where $z$ (typically extracted from discrete meshpoints) is high-dimensional, it is typical that $h(z) \subseteq z$, though exceptions exist, see for instance \cite{eivazi2020deep}. In extended DMD, which is often used in lower-dimensional systems, the lifting $h(z)$ is constructed with a dictionary of lifting functions, such as neural networks \cite{williams2015edmd, brunton2021modern}.

Here, we consider two different sets of observable functions. One maps the observations to themselves: $F=h_f(z)=z$. The resulting state vector $F$ contains the fluid pressure, vertical velocity, and horizontal velocity at every vertex of the simulation mesh. The second observable function $S=h_s(z) \subseteq z$ also contains unlifted states at mesh vertices, but only on the vertices that lie on the surface of $B$, and only contains pressure measurements. In both the free and forced oscillation cases, $B$ is defined as the downstream cylinder. The values of $S$ can then be physically interpreted as measurements from pressure sensors on the surface of the body, which would be feasible to obtain in practical applications, whereas the field values $F$ are likely not possible to construct outside of a controlled laboratory setting. DMD is always performed on a subset of $g(z)$, and though $S$ is a much narrower subset than $F$, modes calculated on either subset are valid DMD modes of the fluid system. This motivates the possibility that the more accessible surface modes may contain information about the broader fluid system that could be useful for flow reconstruction.

\subsection{Numerical DMD}

In this section, we describe our approach to approximate the eigenvalues and eigenvectors of $K$ on $S$ and $F$ from specific windows of time within the simulations. These windows contain $m \in \mathbb{N}$ snapshots, starting from snapshot $b \in \mathbb{N}$, and the window is not allowed to continue past the available data ($m+b \leq M$). The windows of data are taken from a specific simulation, defined by both its type $w \in \{1,2\}$ ($1$ denoting the free oscillations and $2$ denoting the prescribed oscillations), and its specific simulation parameters $d$, with $d=U^* \in \{1,2,\ldots,15\}$ for $w=1$ or $d=A \in \{0.1,0.2,\ldots,1.0\}$ for $w=2$. Let the set $C$ define the set of unique tuples $(m,b,w,d)$ that it is possible to construct under the given constraints. Here we demonstrate the DMD approach for an arbitrary $c \in C$. Because all of the quantities defined beyond this point, such as the data windows, operators, and modes, depend on $c$, we drop this dependence from our notation for conciseness. 

The observations from the simulation can be constructed into time-shifted surface data arrays $X \in \RR^{k_{s} \times (m-1)}$ and $\bar{X} \in \RR^{k_{s} \times (m-1)}$, as well as field data arrays and $Y \in \RR^{k_{f} \times (m-1)}$ and $\bar{Y} \in \RR^{k_{s} \times (m-1)}$ (where $k$ denotes the number of meshpoints and the subscripts $s$ and $f$ generally refer to quantities calculated for the surface data and broader field data, respectively), as

\begin{align}
X &= \begin{bmatrix}
    S_{b} & S_{b+1} & \cdots & S_{b+m-1}
\end{bmatrix}\\
\bar{X} &= \begin{bmatrix}
    S_{b+1} & S_{b+2} & \cdots & S_{b+m}
\end{bmatrix}\\
Y &= \begin{bmatrix}
    F_{b} & F_{b+1} & \cdots & F_{b+m-1}
\end{bmatrix}\\
\bar{Y} &= \begin{bmatrix}
    F_{b+1} & F_{b+2} & \cdots & F_{b+m}
\end{bmatrix}.
\end{align}

The operator on the surface measurements $K_{s} \in \RR^{k_{s} \times k_{s}}$ and the operator of the field measurements $K_{f} \in \RR^{k_{f} \times k_{f}}$ can then be found by recognizing that these matrices can be used to reconstruct the optimization problem in eq. \ref{eqn:minimization} as
\begin{align}
    K_{s}=\argmin_A \|\bar{X} - A X\|_F\\
    K_{f}=\argmin_A \|\bar{Y} - A Y\|_F,
\end{align}
where $\| \cdot \|_F$ denotes the Frobenius norm. The operators resulting from this optimization map the data snapshots forward in time with minimal error in a least-squares sense, for instance: $S_{n+1} \approx K_{s} S_{n}$ and $F_{i+n} \approx K_{f} F_{n}$. If $m-1 \leq k_{f}$, in other words, the data matrices have at least as many rows as they have columns (as is the case here), then the forward prediction has only roundoff error. Our objective is to calculate the eigenvalues and eigenvectors of the $K$ matrices. However, $K$ can be large, and calculating the eigendecomposition of a large matrix is computationally challenging. For instance, the free oscillation case has 32,434 mesh vertices, each with three states, so $K_{f}$ has $k_{f}^2=9.47 \times 10^9$ parameters, which is a challenge to store in random-access memory on most modern computers, and even more challenging to perform computations on. On the fluid field data, this challenge is avoided by first calculating the singular value decomposition (SVD),
\begin{equation}
    Y=U \Sigma V^T,
\end{equation}
where $T$ denotes the transpose, $U \in \RR^{k_{f} \times k_{f}}$ and $V \in \RR^{(m-1) \times (m-1)}$ are unitary matrices and $\Sigma \in \RR^{k_f \times (m-1)}$ is a diagonal matrix containing singular values, by convention sorted in descending order from the top left. These matrices can be truncated to improve the computational efficiency of the following steps. Truncating the matrices such that only the $r$ largest singular values are retained can dramatically reduce the scale of the eigendecomposition with minimal loss of accuracy. Assuming $r \leq max(k,m-1)$, the truncated matrices are denoted $\tilde{\Sigma} \in \RR^{r \times r}$, $\tilde{U} \in \RR^{k_{f} \times r}$ and $\tilde{V} \in \RR^{(m-1) \times r}$. Using the truncated SVD, matrix $K_{f}$ can then be approximated as $\tilde{K}_{f} \in \RR^{r \times r}$ as
\begin{equation}
    \tilde{K}_{f}=\tilde{U}^T \bar{Y} \tilde{V} \tilde{\Sigma}^{-1},
\end{equation}
and its eigenvalues and eigenvectors can be computed as 
\begin{equation}
    \tilde{K}_{f} \Psi_{l}=\Lambda_{f} \Psi_{l},
\end{equation}
where $\Lambda_{f} \in \CC^{r \times r}$ is a diagonal matrix containing the complex eigenvalues
\begin{equation}
    diag(\Lambda_{f})=\begin{bmatrix}
        \lambda_{f,1} & \lambda_{f,2} & \cdots & \lambda_{f,r}
    \end{bmatrix} \in \CC^r
\end{equation}
and the columns of $\Psi_{l} \in \CC^{r \times r}$ contain the eigenvectors in the reduced space, which have little physical meaning at this stage. These eigenvectors can be projected back to the full space using the left singular matrix,
\begin{equation}
    \Psi=\tilde{U} \Psi_{l},
\end{equation}
where the columns of  $\Psi \in \CC^{k_{f} \times r}$ physically correspond to modes of the fluid field
\begin{equation}
    \Psi=\begin{bmatrix}
        \psi_{1} & \psi_{2} & \cdots & \psi_{p}
    \end{bmatrix}.
\end{equation}
Using these modes, it is possible to reconstruct and extrapolate the input data as 
\begin{equation}
    F_{b+q}=\Psi \Lambda_{f}^q \boldsymbol{\alpha}_{f},
    \label{eq:reconstruction}
\end{equation}
where $q \in \mathbb{Z}$ and $\boldsymbol{\alpha}_{f}=\Psi^{-1} F_{b} \in \CC^r$ is a vector of complex numbers
\begin{equation}
    \boldsymbol{\alpha_{f}}=\begin{bmatrix}
        \alpha_{f,1} & \alpha_{f,2} & \cdots & \alpha_{f,r}
    \end{bmatrix} \in \CC^r,
\end{equation}
where $\alpha_{f,j}$ contains the magnitude of mode $\psi_{j}$ at the initial flow snapshot $F_{b}$, and its phase encodes the phase of the mode at the same snapshot. For this reason, we refer to $|\alpha_{f,j}|$ (where $| \cdot |$ denotes the complex magnitude) as the magnitude of mode $j$, and to $\angle \alpha_{f,j}$ (where $\angle \cdot$ denotes the complex phase) as the phase of mode $j$. We will later show that the steady-state flow is often dominated by a small number of high-magnitude modes, which we can exploit to simplify the flow field estimation. 

The number of simulated pressure sensors on the surface is much less than the total number of simulation nodes in the fluid, and only one state (pressure) is measured at each of those surface points (as opposed to both pressure and velocity in the field), which results in $k_{s}<<k_{f}$.  As a result, there is no need to compute the eigendecomposition in a reduced space, and $K_{s}$ can be directly calculated without truncation as
\begin{equation}
K_{s}=\bar{X} X^+,
\label{pseudoinverse}
\end{equation}
where $+$ represents the Moore-Penrose Pseudoinverse. This variation of DMD is known as `Exact DMD' \cite{taira2017modal}, and is equivalent to the SVD-based method without truncation. The eigendecomposition
\begin{equation}
    K_{s} \Phi = \Lambda_{s} \Phi,
\end{equation}
reveals the fluid eigenvectors $\Phi \in \CC^{k_s \times k_s}$ and the eigenvector matrix $\Lambda_{s} \in \CC^{k_s \times k_s}$, which are composed of individual modes and eigenvectors as
\begin{align}
    \Phi&=\begin{bmatrix}
        \phi_{1} & \phi_{2} & \cdots & \phi_{k_{s}}
    \end{bmatrix}\\
       diag(\Lambda_s)&=\begin{bmatrix}
        \lambda_{s,1} & \lambda_{s,2} & \cdots & \lambda_{s,k_{s}}
    \end{bmatrix}.
\end{align}
Much like the broader flow field, the surface pressure field can be reconstructed by a superposition of modes as
\begin{equation}
    S_{b+q}=\Phi \Lambda_{s}^q \boldsymbol{\alpha}_{s},
\end{equation}
where $\boldsymbol{\alpha}_{s}=\Phi^{-1} S_{b} \in \CC^{k_{s}}$ defines the magnitude and phase of the DMD modes of the pressure in the initial time snapshot on the body in a similar manner to how $\boldsymbol{\alpha_{f}}$ defines them for the general flow.

This work aims to construct the global flow field $F$ given the local pressure field $S$. However, mapping directly between these matrices is challenging because of the number of parameters in $F$. We exploit the simplicity of the underlying dynamics of the flow to simplify the problem by instead mapping the most dominant modes in $\Phi$, selected by the magnitude of the corresponding element of $\boldsymbol{\alpha}_{s}$, to the most dominant modes of $\Psi$, determined by the magnitude of the corresponding elements of $\boldsymbol{\alpha}_{f}$. This allows reconstruction of the flow to a reasonable degree of accuracy using only 3 to 4 modes.

\section{Flow Reconstruction}

\subsection{Mode Selection}

For both the free and forced oscillation cases, a strategy must be developed to determine how many modes are necessary to reconstruct the flow and which modes should be used. Let the dominant flow modes which are used in the reconstruction be labeled $[ \tilde{\psi}_{1} \; \tilde{\psi}_{2} \; \cdots \; \tilde{\psi}_{a} ]$, and the most dominant surface modes in the reconstruction be labeled $[ \tilde{\phi}_{1} \; \tilde{\phi}_{2} \; \cdots \; \tilde{\phi}_{a} ]$, where $a \in \mathbb{N}$ is the number of modes selected. Similarly, let the flow field eigenvalue $\tilde{\lambda}_{f,i}$ and magnitude $\tilde{\alpha}_{f,i}$ correspond to the mode $\tilde{\psi}_{i}$, and the surface eigenvalue $\tilde{\lambda}_{s,i}$ and magnitude $\tilde{\alpha}_{s,i}$ correspond to $\tilde{\phi}_{i}$. One key challenge is to make the mapping from surface modes to flow modes consistent for different simulations and different time snapshots within the same simulation. More specifically, for two different windows $c_1$ and $c_2$ in $C$ for the same case $w$ (either free oscillation or forced oscillation), the inner product $<\tilde{\phi}_{i}(c_1),\tilde{\phi}_{i}(c_2)>$ and $<\tilde{\psi}_{i}(c_1),\tilde{\psi}_{i}(c_2)>$ should have magnitude near one and $\tilde{\lambda}_{i}(c_1) \approx \tilde{\lambda}_{i}(c_2)$, implying that the modes are physically similar and correspond to the same physical phenomenon. This is necessary for the neural networks to work well, as the mapping is much simpler when the changes in the modes are small and consistent. By simply ordering the modes based on their relative dominance, this criterion may not necessarily be met: as one mode overtakes another with a change in simulation parameters $U^*$ or $A$, the order of the two modes would flip. To counteract this, a strategy to enforce consistency is necessary, which we tailor for each of the two cases to capture the most important modes throughout the considered ranges of $U^*$ and $A$. 

\subsubsection{Free oscillations in the wake of a stationary obstacle}

\begin{figure}
    \centering
    \captionsetup[subfigure]{labelformat=empty}
    \begin{subfigure}[b]{0.8\hsize}
        \centering
        \includegraphics[width=.98\hsize]{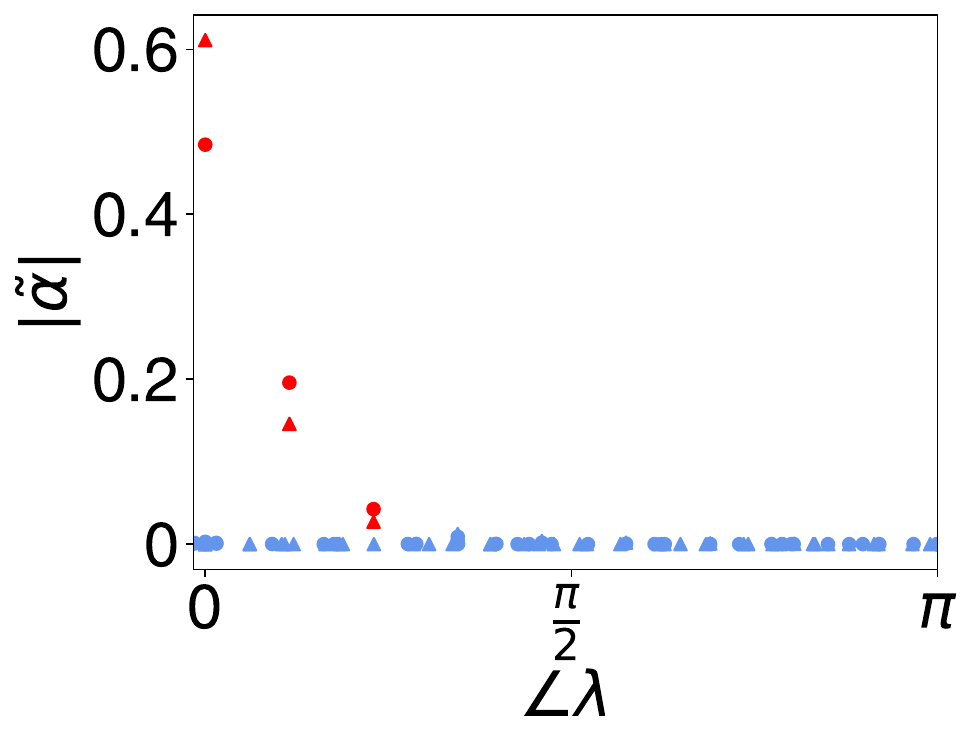}
        \caption{$(a)$}
    \end{subfigure}
    \begin{subfigure}[b]{0.49\hsize}
        \centering
        \includegraphics[width=.98\hsize]{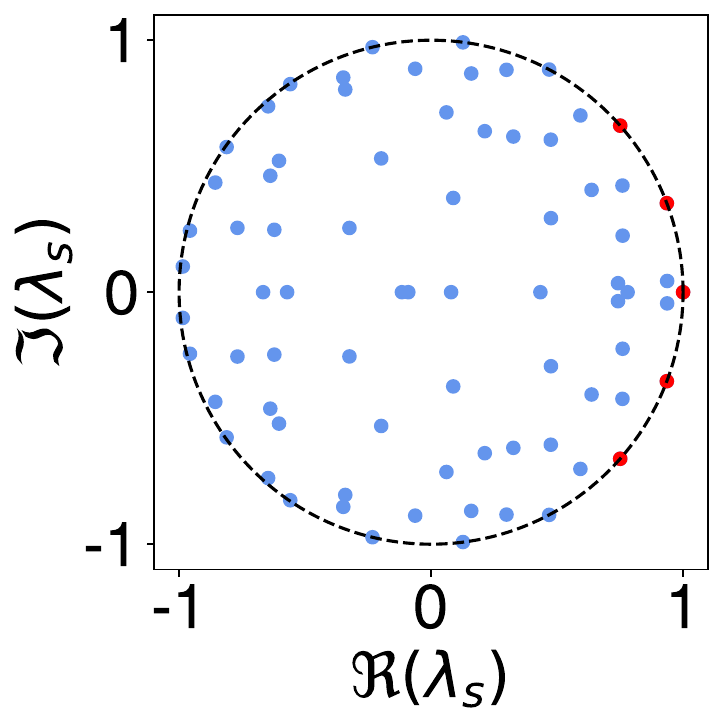}
        \caption{$(b)$}
    \end{subfigure}
    \begin{subfigure}[b]{0.49\hsize}
        \centering
        \includegraphics[width=.98\hsize]{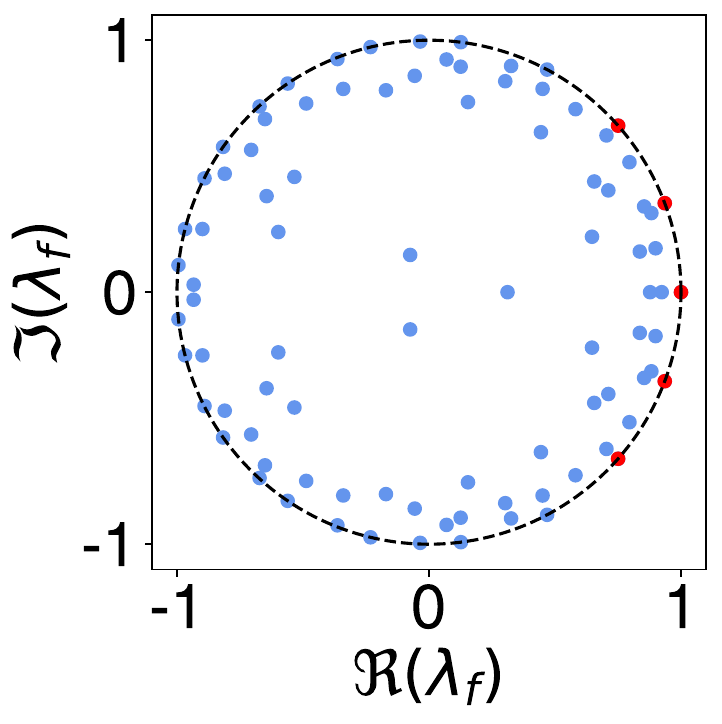}
        \caption{$(c)$}
    \end{subfigure}
    \caption{(a) The normalized magnitude ($\sum_{c \in C}{|\tilde{\alpha}(c)|}=1$) of the DMD modes for the surface pressure measurements (circles) and the flow field state measurements (triangles) for $U^*=8$ and $b=300$. In both cases, three dominant modes (red) can be clearly seen, with the dominant modes of the field and surface having the same frequency. The eigenvalues for (b) the surface data and (c) the flow field data show that the dominant modes have an eigenvalue magnitude of roughly $1$ (meaning that they do not grow or decay with time), and multiple further harmonics can be seen on the unit circle. Decaying modes within the circle are also present but have a low magnitude. The dominant eigenvalues of the flow and surface are clearly similar.}
    \label{fig:mode_energy}
\end{figure}

The relative magnitudes $\boldsymbol{\alpha}$ and eigenvalues $\Lambda$ of the free oscillation simulations with $U^*=8$ after the transient period ($b=300$) are shown in Figure \ref{fig:mode_energy}. Three dominant modes can be clearly seen in Figure \ref{fig:mode_energy}(a) for both the surface and flow field and are marked in red and appear to contain roughly $3/4$ of the total magnitude of all of the modes. The eigenvalues corresponding to the dominant modes of the surface data and those corresponding to the dominant modes of the flow field data have very similar complex phases, indicating that they correspond to phenomena with the same frequency and likely represent the same physical phenomena. These eigenvalues are shown on the unit circle in the complex plane in Figure \ref{fig:mode_energy}(b,c) with the dominant modes shown in red. Many eigenvalues are in the unit circle, corresponding to modes that decay rapidly after the initial transient period. The dominant modes all have complex magnitudes close to $1$ and are on the unit circle. The phases of the eigenvalues of the two dominant modes with non-zero imaginary components can be seen to differ by almost exactly a factor of two, indicating that they are harmonics, with the higher-frequency harmonic having lower magnitude.  This pattern of three dominant harmonic modes, one with zero phase and two that are harmonics, is consistent across values of $U^*$ and across both the surface data and the flow field data. As a result, $a=3$ modes are used for the flow reconstruction for this case. However, certain edge cases can make the mode labeling more challenging for certain windows of data.

The first labeled mode $\tilde{\psi}_{1}$ is always the mode corresponding to the eigenvalue with zero phase, which always has the highest magnitude of ${\alpha}_{f}$. Its index can be found as
\begin{equation}
     i_{1} = \argmax_{i}{ (|\alpha_{f,i}|)}.
\end{equation}
The second mode, corresponding to the first harmonic, is more challenging to identify because it is not always the mode with the second-highest magnitude. Rarely, a mode with very low frequency ($-0.001<\angle \lambda_{f,i}<0.001$) can be found, which can have high magnitude because of overlap with the zero-frequency mode. This low-frequency behavior is not observed in most time windows nor in simulation, so it is considered a numerical artifact and ignored. Accounting for this, the second mode can be identified as
\begin{equation}
    i_2 = \argmax_{i \neq i_{1}} (|\alpha_{f,i}|) \; s.t. \; 0.001<\angle \lambda_{f,i} ,
\end{equation}
which also consistently only finds the positive complex conjugate. To consistently identify the third mode (second harmonic), an additional edge case must be considered: an additional high magnitude mode rarely appears with an eigenvalue phase very near to the phase of the eigenvalue corresponding to $\tilde{\psi}_{2}$, which is also considered a numerical error and neglected. The third mode index is then identified as

\begin{align}
    i_3 = \argmax_{i \notin \{i_{1},i_{2}\}} (|\alpha_{f,i}|) \;
    s.t.  \; 1.1 \angle \lambda_{f,i_2} < \angle \lambda_{f,i},
\end{align}
which eliminates the risk of identifying lower frequency mode as the second harmonic by considering only modes with corresponding frequency at least $1.1$ times greater than the first harmonic. Knowing the indices of the modes, the modes themselves, as well as their eigenvalues and magnitudes, can be defined as $\tilde{\psi}_{j}=\psi_{i_{j}}$, $\tilde{\lambda}_{f,j}=\lambda_{f,i_{j}}$, and $\tilde{\alpha}_{f,j}=\alpha_{f,i_{j}}$ for all $j \in \{1,2,3\}$. The exact same procedure is applied to the surface modes to identify $\tilde{\phi}_{j}$, $\tilde{\lambda}_{s,j}$, and $\tilde{\alpha}_{s,j}$.

\subsubsection{Forced oscillations of tandem cylinders}
\begin{figure}
    \centering
    \captionsetup[subfigure]{labelformat=empty}
    \begin{subfigure}[b]{0.8\hsize}
        \centering
        \includegraphics[width=.98\hsize]{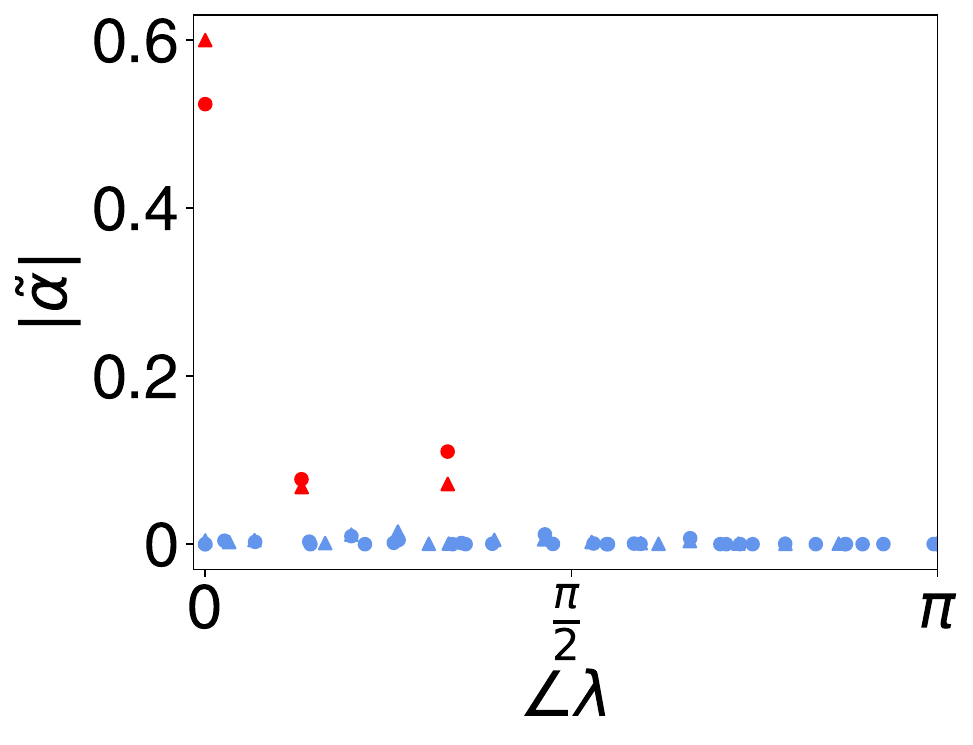}
        \caption{$(a)$}
    \end{subfigure}
    \begin{subfigure}[b]{0.8\hsize}
        \centering
        \includegraphics[width=.98\hsize]{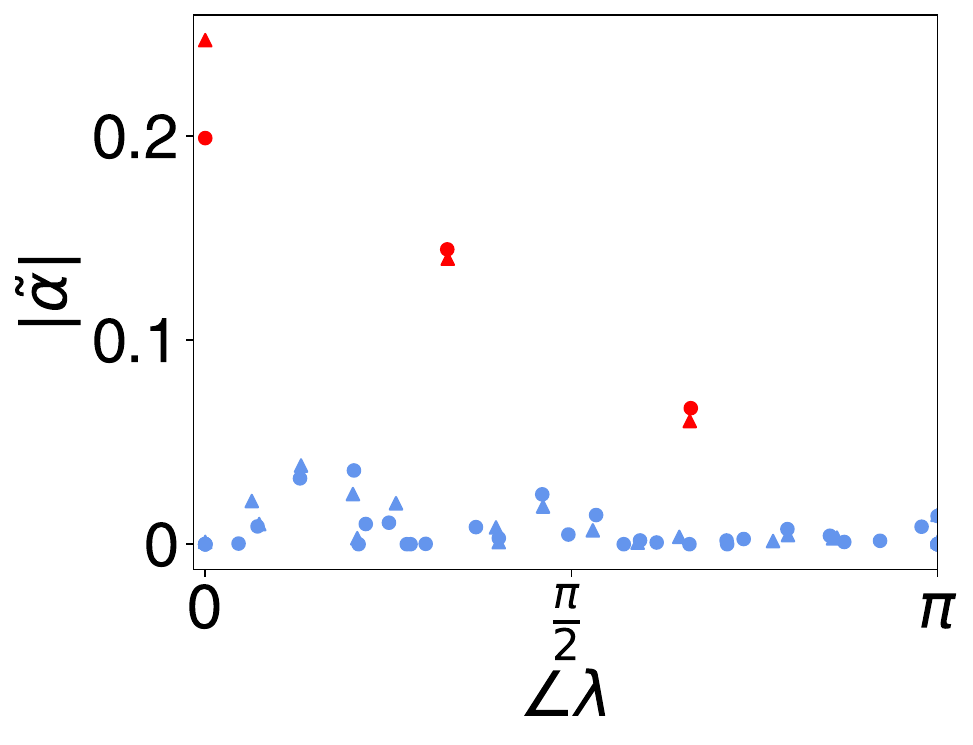}
        \caption{$(b)$}
    \end{subfigure}
    \caption{The relative magnitude of the modes for (a) $A=0.1$ and (b) $A=0.8$ for the surface data (circles) and for the flow field data (triangles). The magnitude is concentrated in fewer modes for the $A=0.1$ case, where three modes can reconstruct the flow with a low degree of lost information. The position of the three most dominant modes (red) changes as a function of $A$, and the magnitude is generally distributed through more modes for higher values of $A$.}
    \label{fig:tandem_energy}
\end{figure}

Identifying the modes in the forced oscillation case is more challenging because the dominant modes change as $A$ varies. For instance, the mode magnitude for $A=0.1$ and $A=0.8$ is shown in Figure \ref{fig:tandem_energy}(a) and (b), respectively. At $A=0.1$, three conjugate pairs of modes have high magnitudes in both the field and on the surface. Similar to the previous case, one has an associated phase of $0$ and is the `mean mode' roughly corresponding to the average value of the measurements. The other two modes have $\angle \lambda_{f,i}$ of $0.41$ and $1.04$. The phase of an eigenvalue can be converted to an angular frequency $\omega$ using the formula
\begin{equation}
    \omega_{i}=\angle \lambda_{f,i} \Delta t,
\end{equation}
where the time between measurements $\Delta t$ is $1$ for this case. The mode with phase $1.04$ rad/s must correspond to a physical phenomenon with frequency $\omega=1.04$ rad/s, the prescribed forcing frequency. This mode, labeled mode 2, corresponds with the periodic fluid behavior driven by these oscillations. The other dominant mode, which we label mode 3, has a frequency $\omega=0.41$, which does not correspond to an integer multiple of the forcing and likely corresponds to the frequency of vortex shedding for the unforced system.

A larger number of modes with significant amplitude can be seen at the higher forcing amplitude of $A=0.8$, shown in \ref{fig:tandem_energy}(b). The three highest magnitude modes exist in descending order of magnitude at $\angle \lambda_{f}$ of $0$, $1.04$, and $2.08$. The former two modes can be identified by their frequencies as modes 1 and 2, respectively, which were identified for the lower forcing amplitude. The mode with $\angle \lambda_{f}=2.08$ had negligible amplitude at the lower forcing amplitude, and based on its frequency, it is the second harmonic of the prescribed forcing. We label this mode as mode 4. Though its magnitude relative to the forcing modes has decreased, mode 3 can also be observed here with a phase of $0.41$.

To estimate these modes, a strategy must again be developed that can identify physically consistent modes of the operators, irrespective of the parameter $A$. A reasonable strategy is to identify them by the phases of their associated eigenvalues as
\begin{align}
    i_1&= \argmax_{i} (|\alpha_{f,i}|) \;s.t. \; \angle \lambda_{f,i}=0 \\
    i_2&= \argmax_{i} (|\alpha_{f,i}|) \;s.t. \; 1.00 < \angle \lambda_{f,i} < 1.08 \\
    i_3&= \argmax_{i} (|\alpha_{f,i}|) \;s.t. \; 0.33 < \angle \lambda_{f,i} < 0.43 \\
    i_4&= \argmax_{i} (|\alpha_{f,i}|) \;s.t. \; 2.00 < \angle \lambda_{f,i} < 2.16.
\end{align}
Because of the possibility of numerical error causing the frequencies of modes 2 and 4 to vary from the known forcing frequency, those modes are identified from a small range centered on the known forcing frequency. The frequency of the vortex shedding that mode 3 corresponds to can vary as a function of $A$, so the range in which to search for mode 3 is determined by identifying the phase of mode 3 for a range of $A$ values. The range of identified values is $0.35 \leq \lambda_{f,i_3} \leq 0.41$, so the range where mode 3 is searched for is the same plus a margin of $0.02$. 

Based on these indices, the modes themselves, as well as their eigenvalues and magnitudes, can be defined in the same manner as for the free oscillation case: $\tilde{\psi}_{j}=\psi_{i_{j}}$, $\tilde{\lambda}_{f,j}=\lambda_{f,i_{j}}$, and $\tilde{\alpha}_{f,j}=\alpha_{f,i_{j}}$. The exact same procedure is again applied to the surface modes to identify $\tilde{\phi}_{j}$, $\tilde{\lambda}_{s,j}$, and $\tilde{\alpha}_{s,j}$. The key difference is that here, $j \in \{1,2,3,4\}$.

\subsection{Data Normalization}

To reconstruct the full flow field based on surface measurements, we need to estimate the values of $\tilde{\psi}_{i}$, $\tilde{\lambda}_{f,i}$, and $\tilde{\alpha}_{f,i}$ given $\tilde{\phi}_{i}$, $\tilde{\lambda}_{s,i}$, and $\tilde{\alpha}_{s,i}$. However, performing this mapping directly using a neural network is challenging, as the input to a traditional neural network must be a vector of real-valued features, and the values to be mapped are complex-valued. Additionally, because modes are eigenvectors, they can be multiplied by any non-zero complex number and remain a valid mode: the result of $v  \tilde{\psi}_{i}$ for $v \in \CC$ where $v \neq 0$ is another valid representation of mode $i$. During the eigendecomposition, one of these valid representations is calculated for $\lambda_{f,i}$, but it is not necessarily done in a consistent manner, so two similar modes $\psi_{i}(c_1)$ and $\psi_{i}(c_2)$ for similar data windows $c_1,c_2 \in C$ could appear quite different only due to having different arbitrary constants. The effect of the complex constant can be decomposed into two parts: its magnitude scales the magnitude of the mode vector $\psi_{i}$, and its phase rotates the phase of every element of $\psi_{i}$ in the complex plane. As a result of this rotation, the absolute phase of the elements of $\psi_{i}$ holds no meaning; however, the relative phases of its elements do hold useful information about the relative timing of state oscillations at different points in the flow field. 

The process of normalizing the phase of a complex-valued vector is less standardized than other forms of normalization. However, the prevalent approach involves phase rotation such that a designated reference element, often the first element in the vector, aligns with the real axis. This rotation preserves the relative phases between the points, thereby facilitating any subsequent analyses that may use that information. Letting $\tilde{\psi}_{i,1}$ denote the first element of $\tilde{\psi}_{i}$, the phase normalized vector $\hat{\psi}_{i}$ can be defined as
\begin{equation}
    \hat{\psi}_{i}=\frac{\tilde{\psi}_{i,1}'}{|\tilde{\psi}_{i,1}|} \tilde{\psi}_{i},
\end{equation}
where $'$ denotes the complex conjugate. Because the values of $\tilde{\alpha}_{f,i}$ were constructed using the pre-normalization modes, their complex phases must be updated (in the opposite direction) as
\begin{equation}
    \hat{\alpha}_{f,i}=\frac{\tilde{\psi}_{i,1}}{|\tilde{\psi}_{i,1}|}\tilde{\alpha}_{f,i}.
\end{equation}
This normalization procedure is also applied to the surface measurements to calculate $\hat{\phi}_i$ and $\hat{\alpha}_{f,i}$.

\subsection{Map Definition}

In order to reconstruct the flow field, we require a map
\begin{align}
    f_c: (\hat{\phi}_{i},\tilde{\lambda}_{s,i},\hat{\alpha}_{s,i}) \to (\hat{\psi}_{i},\tilde{\lambda}_{f,i},\hat{\alpha}_{f,i}).
\end{align}
However, the mapping is performed by a dense neural network, which maps a real-valued vector to another real-valued vector. 
These modes must then be concatenated into a real-valued vector in order to be mapped by the neural network, which is achieved by concatenating its real and imaginary components, as well as the real and imaginary components of its corresponding magnitude,
\begin{align}
    x_{i}=\begin{cases}
        \begin{bmatrix}
        \Re(\hat{\phi}_{i}) & \Re(\hat{\alpha}_{s,i}) 
    \end{bmatrix} & i = 1 \\
    \begin{bmatrix}
        \Re(\hat{\phi}_{i}) &  \Im(\hat{\phi}_{i}) & \Re(\hat{\alpha}_{s,i}) & \Im(\hat{\alpha}_{s,i})
    \end{bmatrix} & i \neq 1
    \end{cases} \\
    y_{i}=\begin{cases}
        \begin{bmatrix}
        \Re(\hat{\psi}_{i}) & \Re(\hat{\alpha}_{f,i}) 
    \end{bmatrix} & i = 1 \\
    \begin{bmatrix}
        \Re(\hat{\psi}_{i}) &  \Im(\hat{\psi}_{i}) & \Re(\hat{\alpha}_{f,i}) & \Im(\hat{\alpha}_{f,i})
    \end{bmatrix} & i \neq 1.
    \end{cases}
\end{align}
When $i=1$, both the mode and magnitude are real, so there is no need to store the imaginary components. The eigenvalues are conspicuously missing from these vectors; that is because their magnitudes are roughly equal because they must be near 1 in the steady state ($|\tilde{\lambda}_{f,i}| \approx |\tilde{\lambda}_{s,i}| \approx 1$), and their phases are always very similar because they describe the same physical phenomena ($\angle \tilde{\lambda}_{f,i} \approx \angle \tilde{\lambda}_{s,i}$), so the eigenvalues are mapped separately by
\begin{equation}
    f_I: \tilde{\lambda}_{s,i} \to \tilde{\lambda}_{f,i},
\end{equation}
where $f_I$ is the identity map. By mapping the eigenvalues separately from the other flow information in this way, the number of parameters in the neural network can be slightly reduced.
The vectors are further concatenated into a longer vectors $\mathbb{X} \in \RR^{(2a-1)(k_{s}+1)}$ and $\mathbb{Y} \in \RR^{(2a-1)(k_{f}+1)}$ that contain information about all of the modes as
\begin{align}
\mathbb{X} =\begin{bmatrix}
    x_{1} & x_{2} & \cdots & x_{a}
\end{bmatrix} \\
\mathbb{Y} =\begin{bmatrix}
    y_{1} & y_{2} & \cdots & y_{a}
\end{bmatrix}.
\end{align}
The number of parameters in $\mathbb{X}$ and $\mathbb{Y}$ is much less than that of $X$ and $Y$ for large $m$, which makes constructing a mapping between them easier. The mapping is performed by a fully-connected dense neural network $\mathcal{N}$ as $\tilde{\mathbb{Y}}=\mathcal{N}(\mathbb{X})$, where $\tilde{\mathbb{Y}}$ is the predicted value of $\mathbb{Y}$. 

The network architecture contains three hidden layers of 500, 1000, and 1500 nodes (in order from input to output), with hyperbolic tangent rectification on the hidden layers and a linear activation function on the output. More specifically, the mapping takes the form
\begin{equation}
    \mathcal{N}(x) = W_4 \tanh(W_3 \tanh (W_2 \tanh (W_1 x+b_1)+b_2)+b_3)+b_4,
\end{equation}
where the weights $W_1 \in \RR^{500 \times (2a-1)(k_{s}+1)}$, $W_2 \in \RR^{1000 \times 500}$, $W_3 \in \RR^{1500 \times 1000}$, $W_4 \in \RR^{(2a-1)(k_f+1) \times 1500}$, $b_1 \in \RR^{500}$, $b_2 \in \RR^{1000}$, $b_3 \in \RR^{1500}$, and $b_4 \in \RR^{(2a-1)(k_{f}+1)}$. Collectively, we refer to the list containing all of these weights as $\sigma$. Each of the two simulation cases requires its own weights, so the weights corresponding to the free oscillations are labeled $\sigma_1$, and the weights corresponding to the prescribed oscillations are labeled $\sigma_2$.
The weights $\sigma_j$ are trained to minimize the loss
\begin{equation}
    \sigma_j=\argmin_{\sigma} \sum_{c \in C_j} \|\mathcal{N_{\sigma}}(\mathbb{X}(c))-\mathbb{Y}(c)\|_2,
    \label{eq:minimization}
\end{equation}
where $C_j$ denotes the set of possible time windows over all simulations for case $j$. This minimizes the error between the expected field mode vector and the true one in a least-squares sense. 

\subsection{Map Implementation}

Computational limitations make it challenging to perform this optimization over the entire set $C_j$, and doing so would leave no new data on which to test the effectiveness of the network. To solve both of these problems, we split $C_j$ into three different sets: a training set $C_{j,t}$, a validation set $C_{j,v}$, and a testing set $C_{j,e}$. The training set contains either points from the free oscillation case where $U^* \in \{1, 2, \ldots, 14\}$ and $b=\{210, 211, \ldots, 310\}$, or points from the forced oscillation case where $A \in \{0.1,0.2, \ldots ,1.0\}$ where $A \neq 5$, with $b=\{50,51,\ldots,90\}$. The validation set includes the same ranges of $U^*$ or $A$, but has $b=\{330,331,\ldots,340\}$ for the free oscillation case or $b=\{140,141,\ldots,149\}$ for the forced oscillation case. This choice of initial times allows the network to be validated with little data overlap. The test data is from simulations not seen in the other datasets: in the free oscillation case, $U^*=15$ is used to demonstrate that this procedure can extrapolate beyond the training parameter range, and in the forced oscillation case, $A=0.5$ to demonstrate interpolation within the training parameter range. The entire steady state time window is used, $b=\{210,211,\ldots,340\}$ for the free oscillation case and $t_0=\{50,51,\ldots,149\}$ for the forced oscillation case. In the free oscillation case $m=100$, and in the forced oscillation case $m=50$.

At the beginning of training, $20$ snapshots from each simulation in $C_{j,t}$ are randomly selected to form a list $\boldsymbol{c}_{j,t} \subseteq C_{j,t}$, and a list of training matrices $\bar{\mathbb{X}}_t=[\mathbb{X}(\boldsymbol{c}_{j,t,1}), \mathbb{X}(\boldsymbol{c}_{j,t,2}), \ldots, \mathbb{X}(\boldsymbol{c}_{j,t,20})]$ and $\bar{\mathbb{Y}}_t=[\mathbb{Y}(\boldsymbol{c}_{j,t,1}), \mathbb{Y}(\boldsymbol{c}_{j,t,2}), \ldots, \mathbb{Y}(\boldsymbol{c}_{j,t,20})]$. Validation data $\bar{\mathbb{X}}_v$ and $\bar{\mathbb{Y}}_v$ are also constructed by the same procedure, however, using only $5$ snapshots per simulation. This data is iteratively used to improve $\mathcal{N}_{\sigma}$ by minimizing its mean-squared training loss, given by a slightly modified version of eq. \ref{eq:minimization}:
\begin{equation}
    L_{t,j} = \sum_{c \in \boldsymbol{c}_{j,t}} \|\mathcal{N_{\sigma_j}}(\mathbb{X}(c))-\mathbb{Y}(c)\|_2,
    \label{eq:minimization_loss}
\end{equation}
where validation loss $L_{v,j}$ is calculated by the same procedure over the validation data. Because all of the losses and weights depend on the case $j$, we drop it from the subscripts beyond this point for clarity. This loss is then iteratively minimized using the \emph{adam} algorithm, which is a stochastic gradient descent algorithm with momentum. The gradient of $L_{t}$ with respect to $\sigma$ is calculated on a subset (batch) of the training data $\boldsymbol{C}_{t}$, and the weights are updated in the direction of decreasing loss with an additional `momentum' term based on the gradients of previous batches. The batches are iterated until all of the data has been used, known as an \textit{epoch}. 

\begin{figure*}[hbt!]
    \centering
    \captionsetup[subfigure]{labelformat=empty}
    \begin{subfigure}[b]{0.53\textwidth}
        \centering
        \includegraphics[width=.98\hsize]{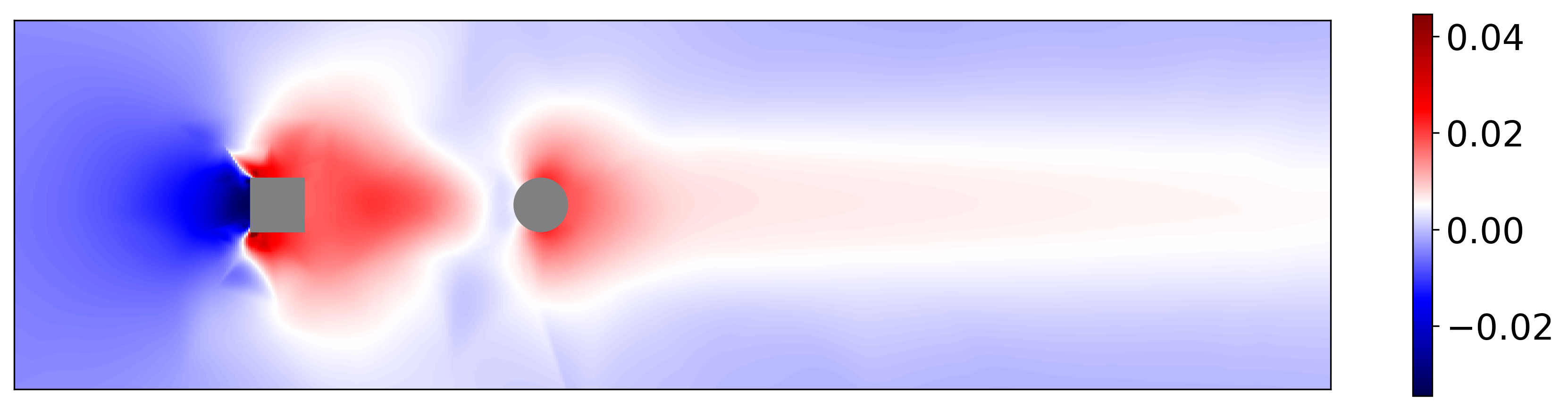}
        \caption{$(a)$ Actual $\Re(\tilde{\psi}_{1})$}
    \end{subfigure}
    \begin{subfigure}[b]{0.46\textwidth}
        \centering
        \includegraphics[width=.98\hsize]{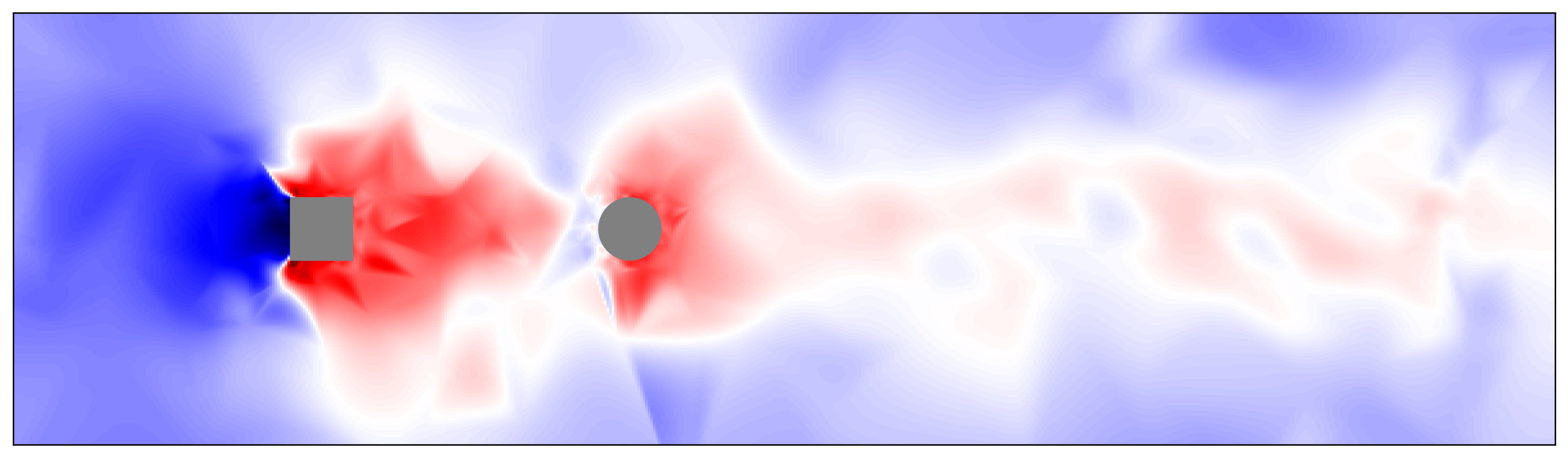}
        \caption{$(b)$ Estimated $\Re(\tilde{\psi}_{1})$}
    \end{subfigure}
    \begin{subfigure}[b]{0.53\textwidth}
        \centering
        \includegraphics[width=.98\hsize]{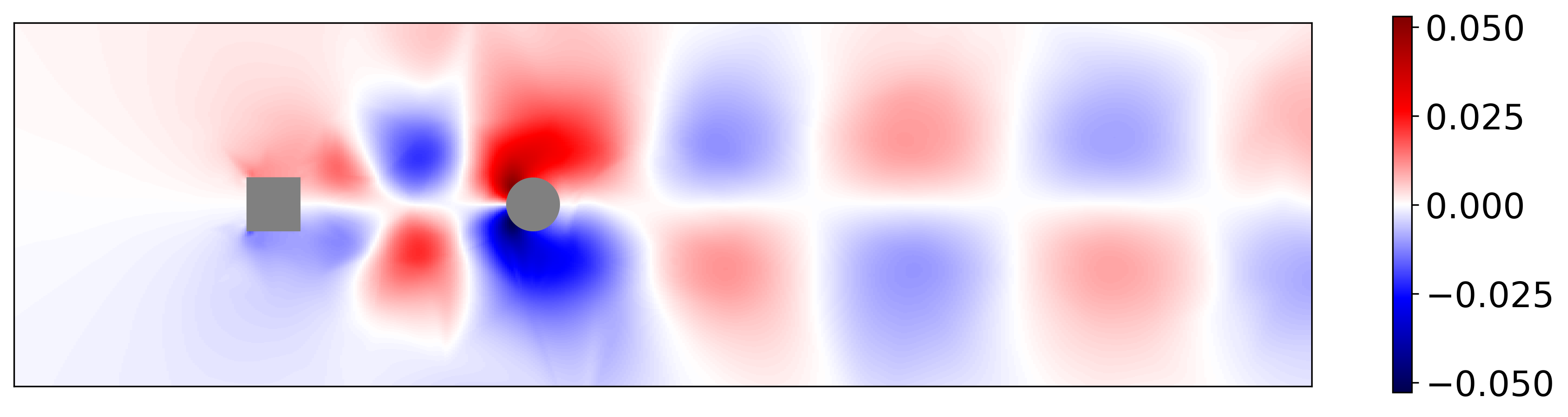}
        \caption{$(c)$ Actual $\Re(\tilde{\psi}_{2})$}
    \end{subfigure}
    \begin{subfigure}[b]{0.46\textwidth}
        \centering
        \includegraphics[width=.98\hsize]{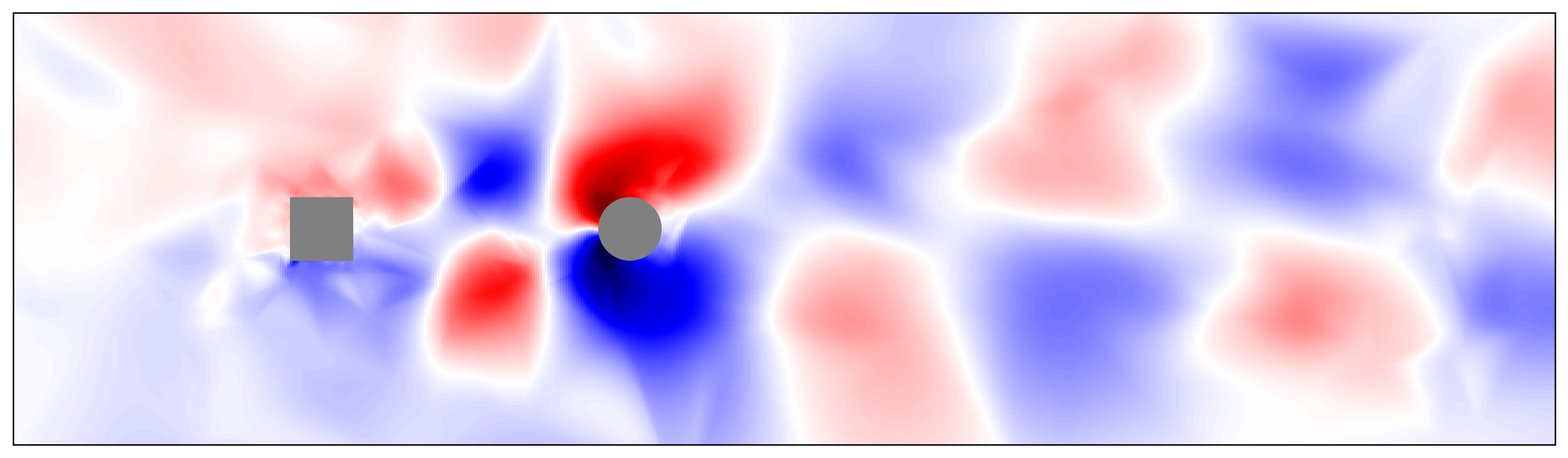}
        \caption{$(d)$ Estimated $\Re(\tilde{\psi}_{2})$}
    \end{subfigure}
    \begin{subfigure}[b]{0.53\textwidth}
        \centering
        \includegraphics[width=.98\hsize]{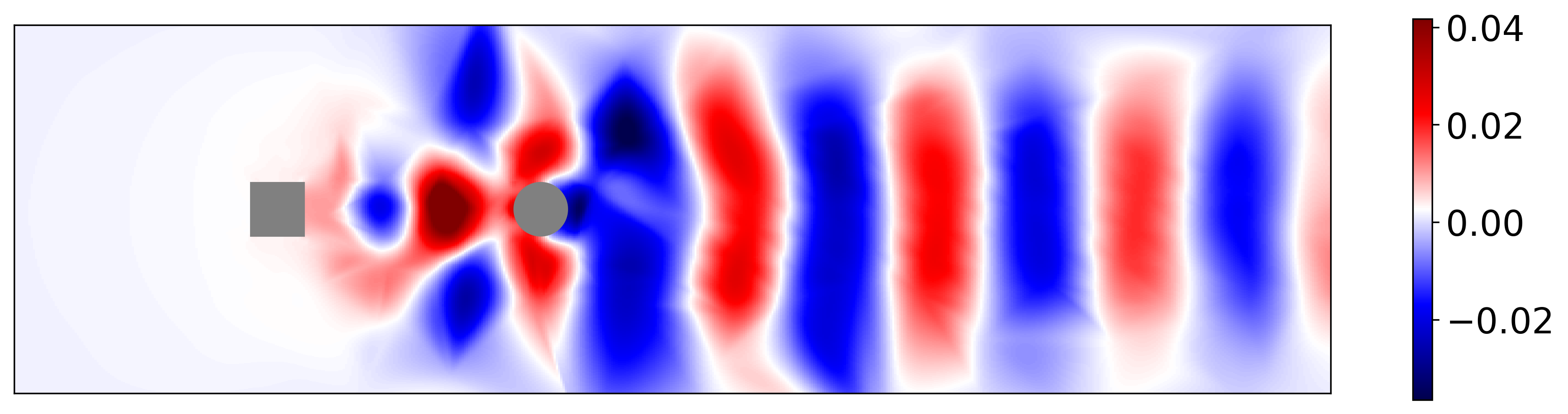}
        \caption{$(e)$ Actual $\Re(\tilde{\psi}_{3})$}
    \end{subfigure}
    \begin{subfigure}[b]{0.46\textwidth}
        \centering
        \includegraphics[width=.98\hsize]{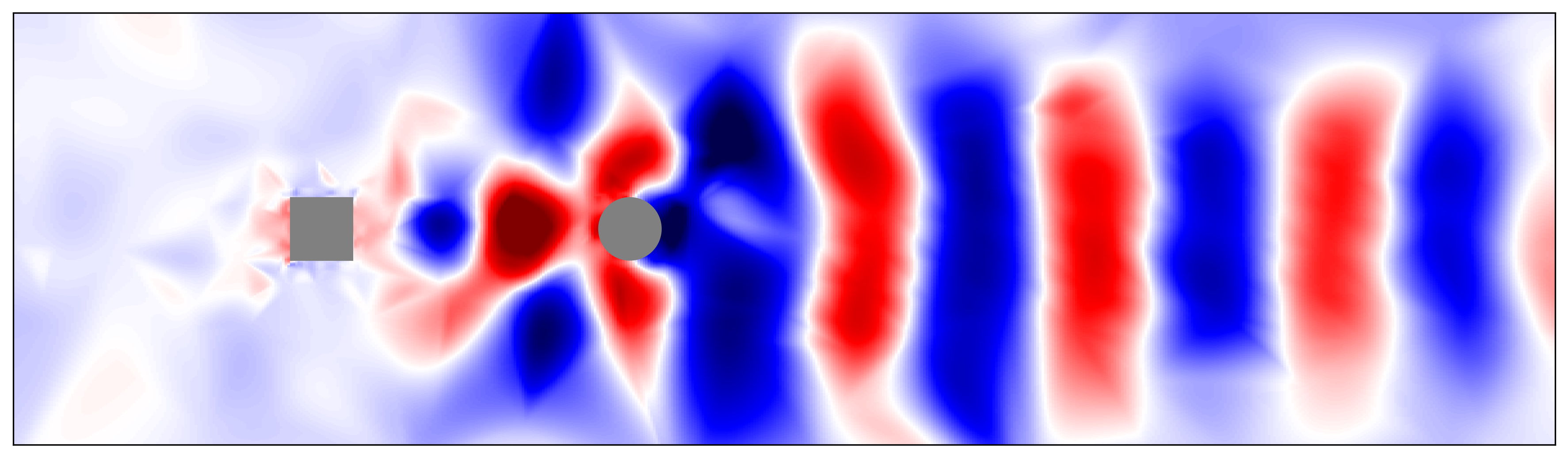}
        \caption{$(f)$ Estimated $\Re(\tilde{\psi}_{3})$}
    \end{subfigure}
    \caption{The real component of the actual vs. estimated modes of the pressure field for free oscillations at $U^*=15$. The estimated modes are qualitatively similar to the actual modes, though the estimation mapping does introduce noise especially close to the leading body. }
    \label{fig:mapped_modes_viv}
\end{figure*}

\begin{figure*}[hbt!]
    \centering
    \captionsetup[subfigure]{labelformat=empty}
    \begin{subfigure}[b]{0.53\textwidth}
        \centering
        \includegraphics[width=.98\hsize]{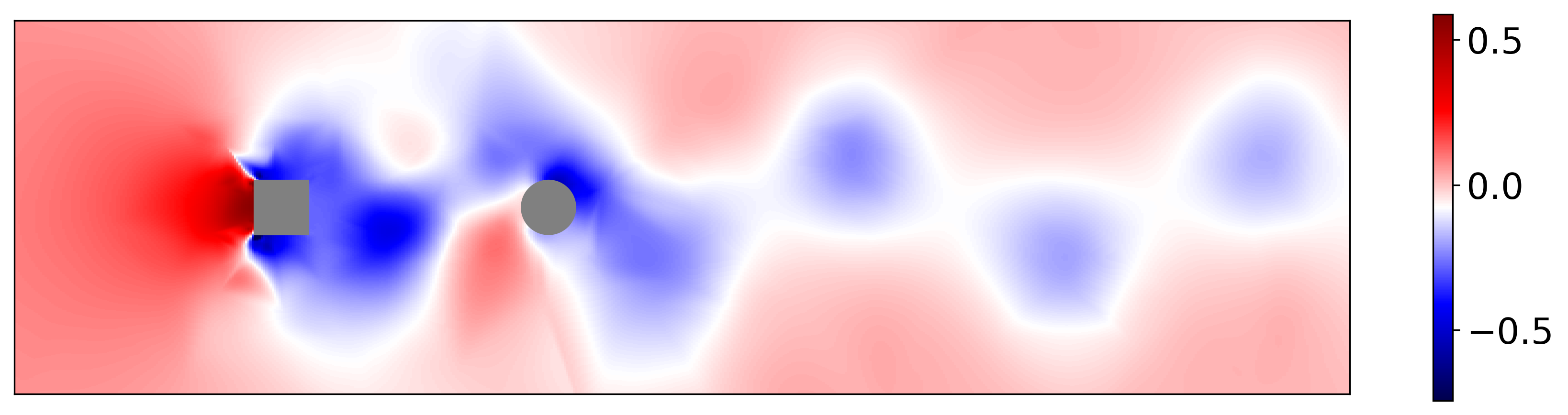}
        \caption{$(a)$ Actual pressure field at $t_0$}
    \end{subfigure}
        \begin{subfigure}[b]{0.46\textwidth}
        \centering
        \includegraphics[width=.98\hsize]{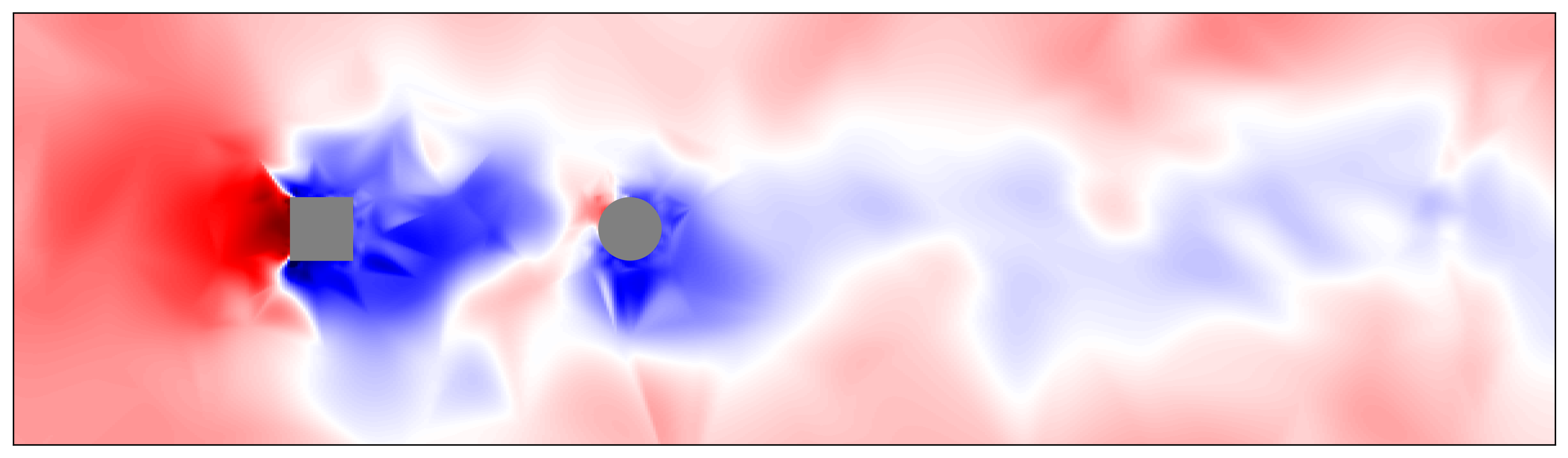}
        \caption{$(b)$ Reconstructed pressure field at $t_0$}
    \end{subfigure}
    \begin{subfigure}[b]{0.53\textwidth}
        \centering
        \includegraphics[width=.98\hsize]{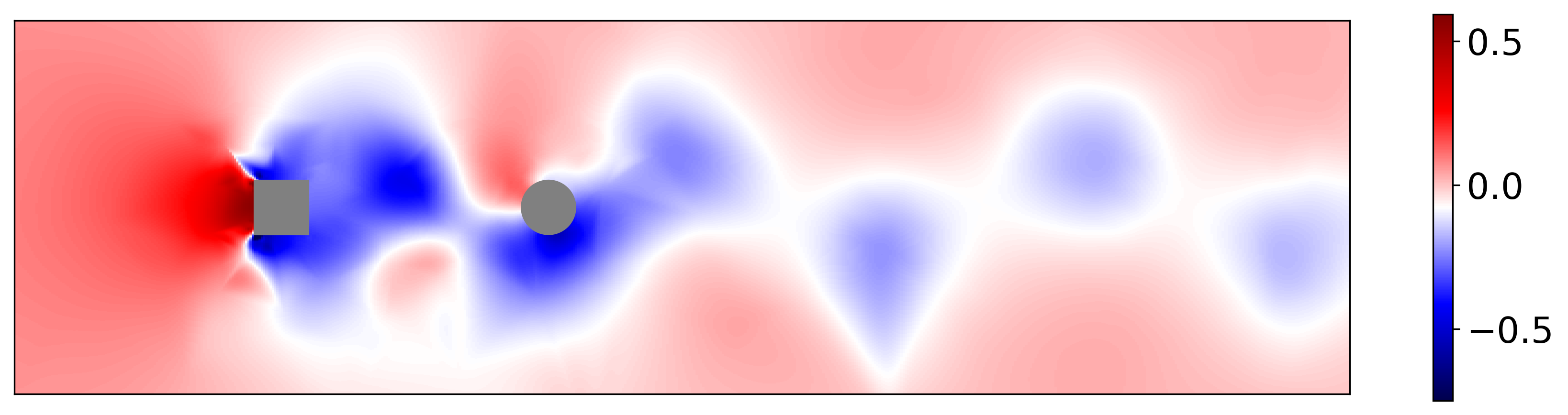}
        \caption{$(c)$ Actual pressure field at $t_0+10$}
    \end{subfigure}
        \begin{subfigure}[b]{0.46\textwidth}
        \centering
        \includegraphics[width=.98\hsize]{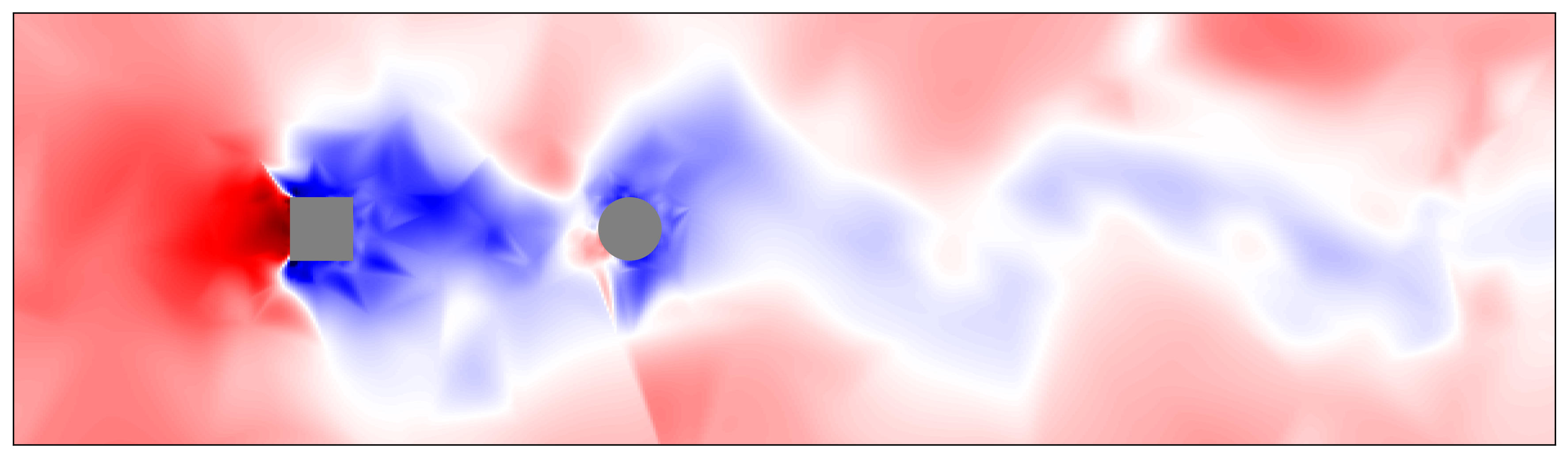}
        \caption{$(d)$ Reconstructed pressure field at $t_0+10$}
    \end{subfigure}
    \begin{subfigure}[b]{0.53\textwidth}
        \centering
        \includegraphics[width=.98\hsize]{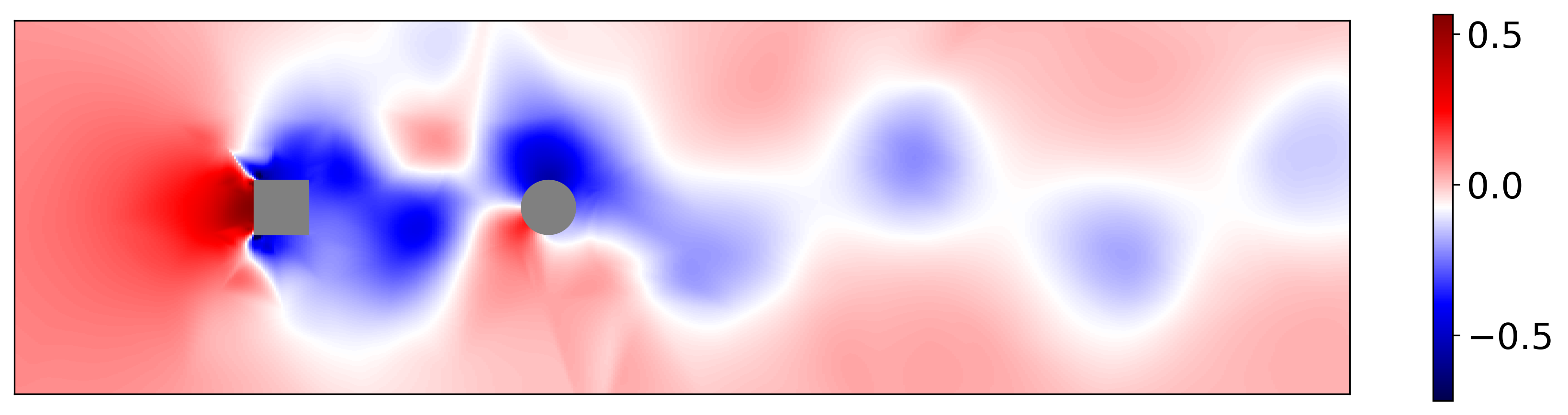}
        \caption{$(e)$ Actual pressure field at $t_0+20$}
    \end{subfigure}
        \begin{subfigure}[b]{0.46\textwidth}
        \centering
        \includegraphics[width=.98\hsize]{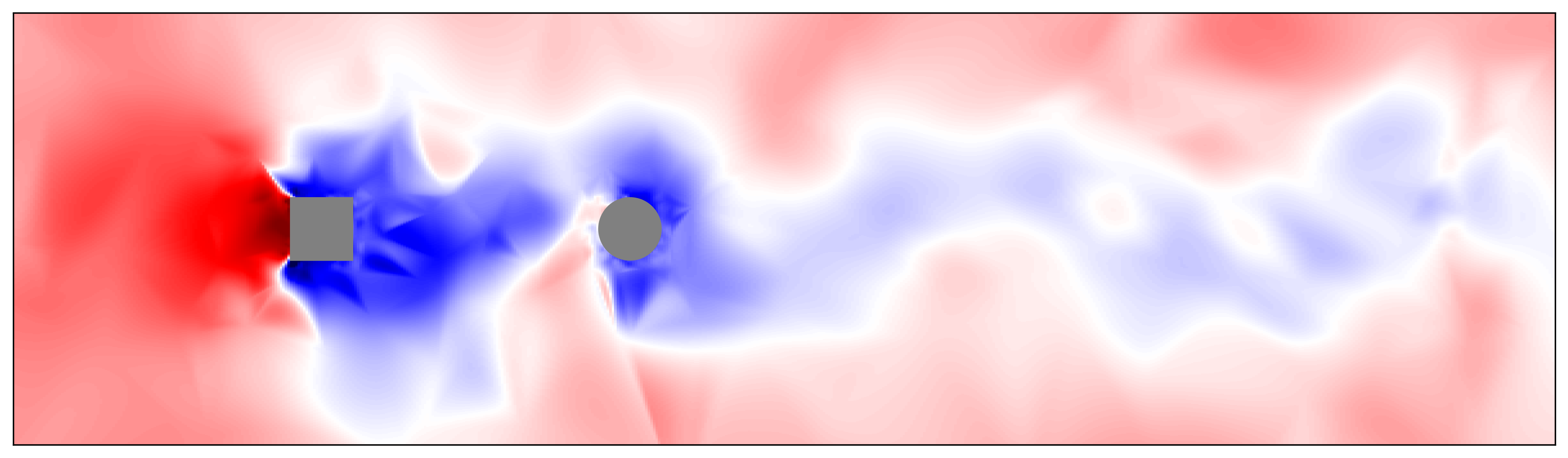}
        \caption{$(f)$ Reconstructed pressure field at $t_0+20$}
    \end{subfigure} 
    \caption{The actual vs. reconstructed pressure field for the case of the wake induced free oscillation of the cylinder  at $U^*=15$ at times $t_0$ (a,b) , $t_0+10$ (c,d), and $t_0+20$ (e,f). } 
    \label{fig:reconstructed_pressure_case1}
\end{figure*}

We use an early stopping algorithm from \cite{prechelt1998early} to know when to terminate the training before overtraining can occur. After every 10\textsuperscript{th} epoch, the loss $L_{v}$ on the validation data is calculated. If $L_{v}<L_{min}$, where $L_{min}$ is the lowest validation error yet recorded, then $L_{min}:=L_v$ and $\sigma_{opt}:=\sigma$, where $\sigma_{opt}$ are the weighs corresponding to $L_{min}$. However, if $L_v>1.2 L_{min}$, and at least 100 epochs have passed, then it is assessed that the network is overtrained and the training is terminated. The weights $\sigma_{opt}$ are then used to reconstruct the flow for case $j$.

\begin{figure*}[ht]
    \centering
    \captionsetup[subfigure]{labelformat=empty}
    \begin{subfigure}[b]{0.53\textwidth}
        \centering
        \includegraphics[width=.98\hsize]{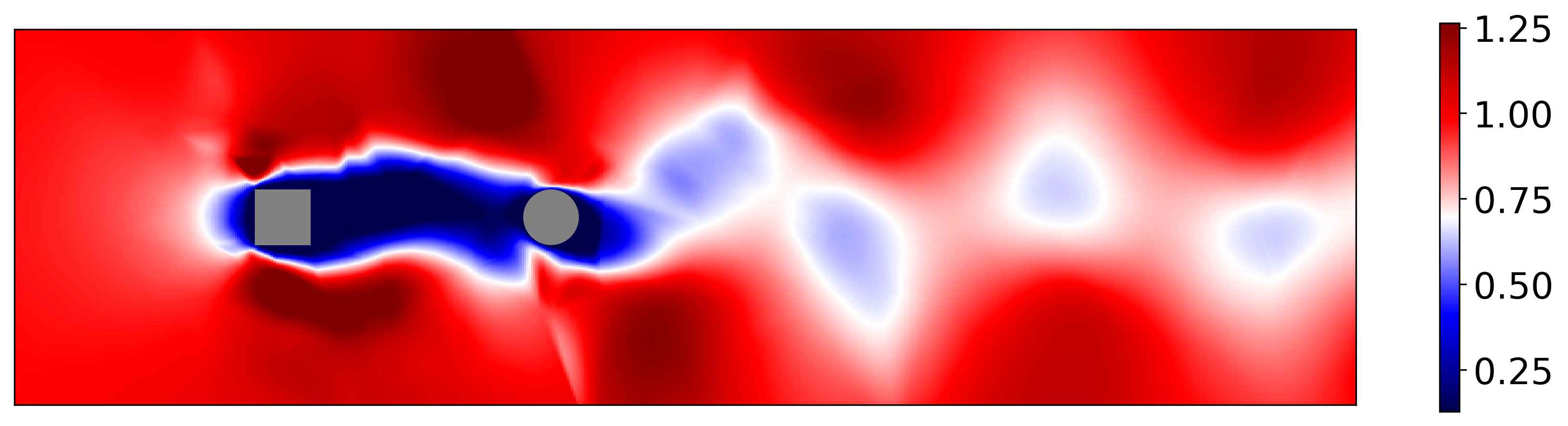}
        \caption{$(a)$ Actual $u_x$ at $t_0$}
    \end{subfigure}
        \begin{subfigure}[b]{0.46\textwidth}
        \centering
        \includegraphics[width=.98\hsize]{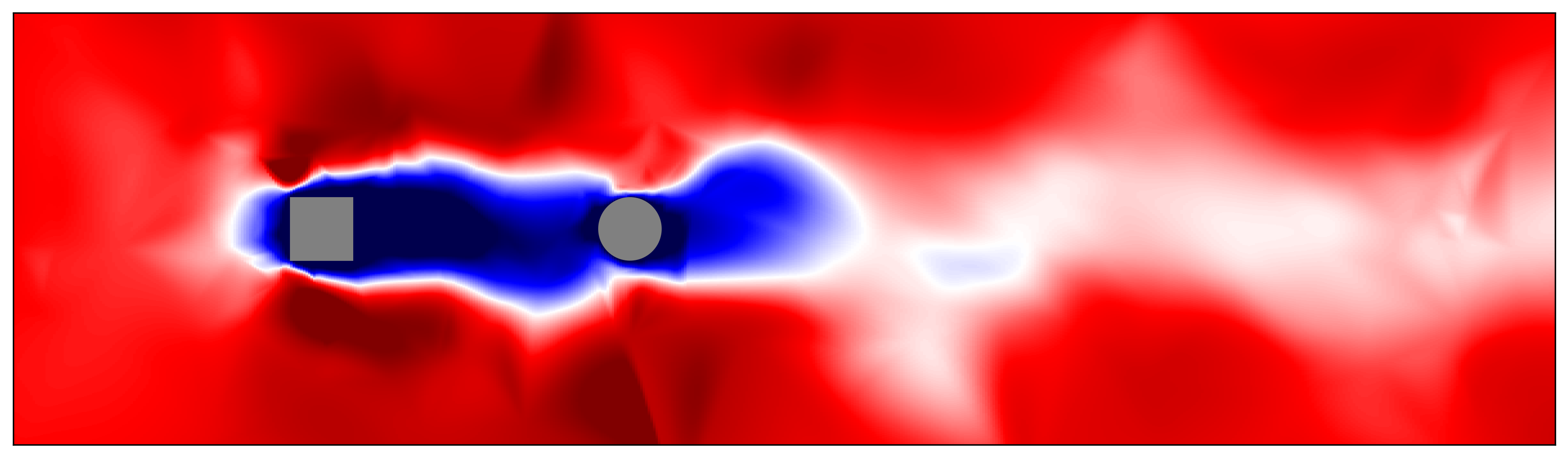}
        \caption{$(b)$ Reconstructed $u_x$ at $t_0$}
    \end{subfigure}
    \begin{subfigure}[b]{0.53\textwidth}
        \centering
        \includegraphics[width=.98\hsize]{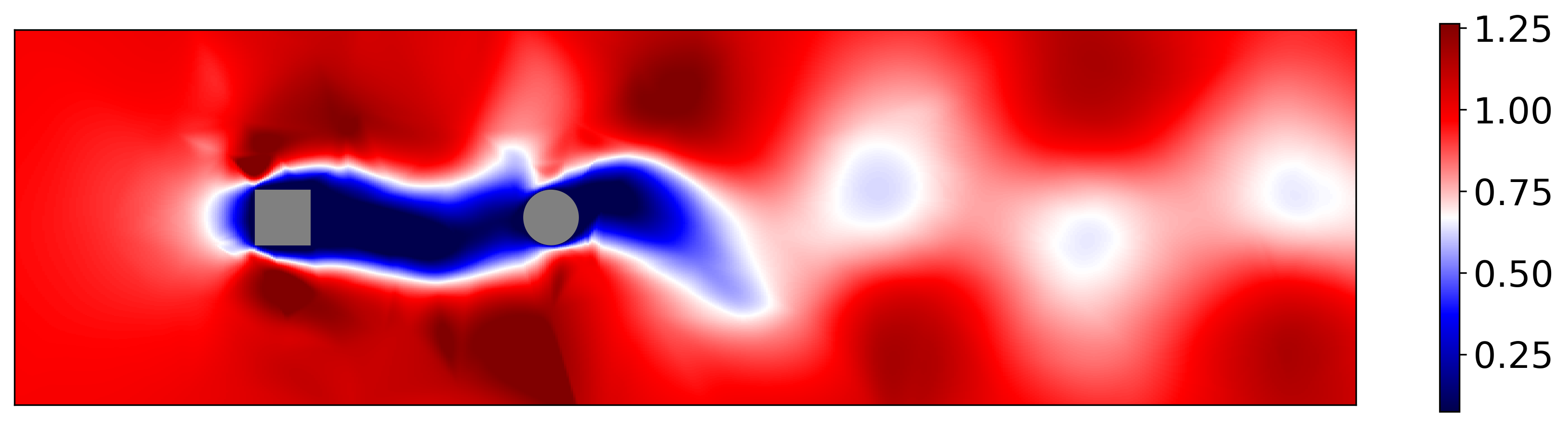}
        \caption{$(c)$ Actual $u_x$ at $t_0+10$}
    \end{subfigure}
        \begin{subfigure}[b]{0.46\textwidth}
        \centering
        \includegraphics[width=.98\hsize]{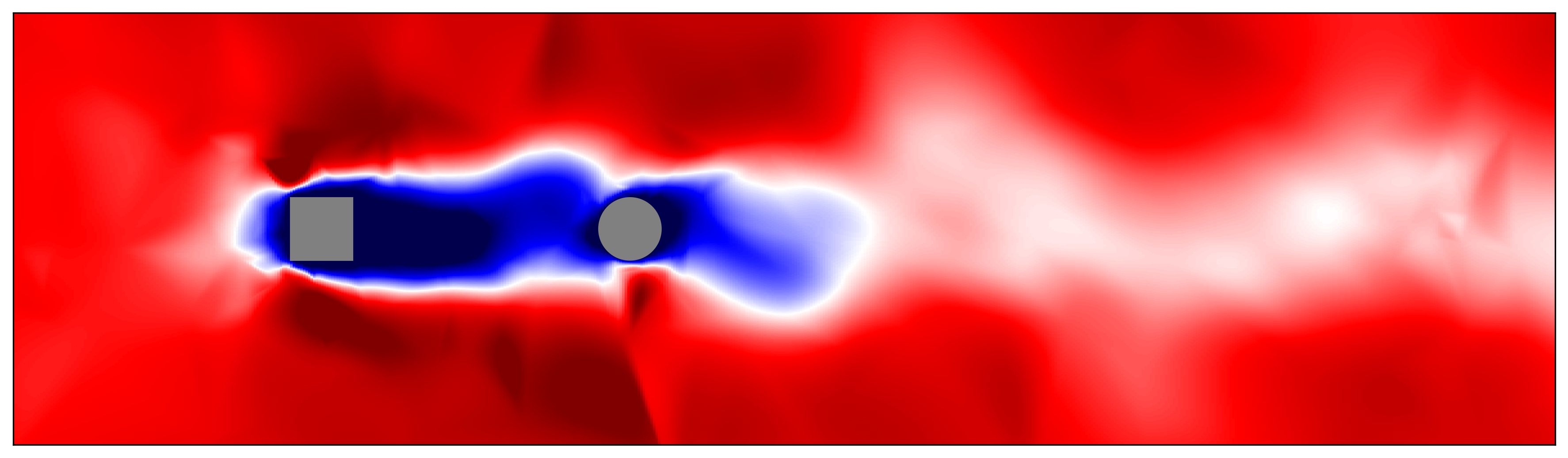}
        \caption{$(d)$ Reconstructed $u_x$ at $t_0+10$}
    \end{subfigure}
    \begin{subfigure}[b]{0.53\textwidth}
        \centering
        \includegraphics[width=.98\hsize]{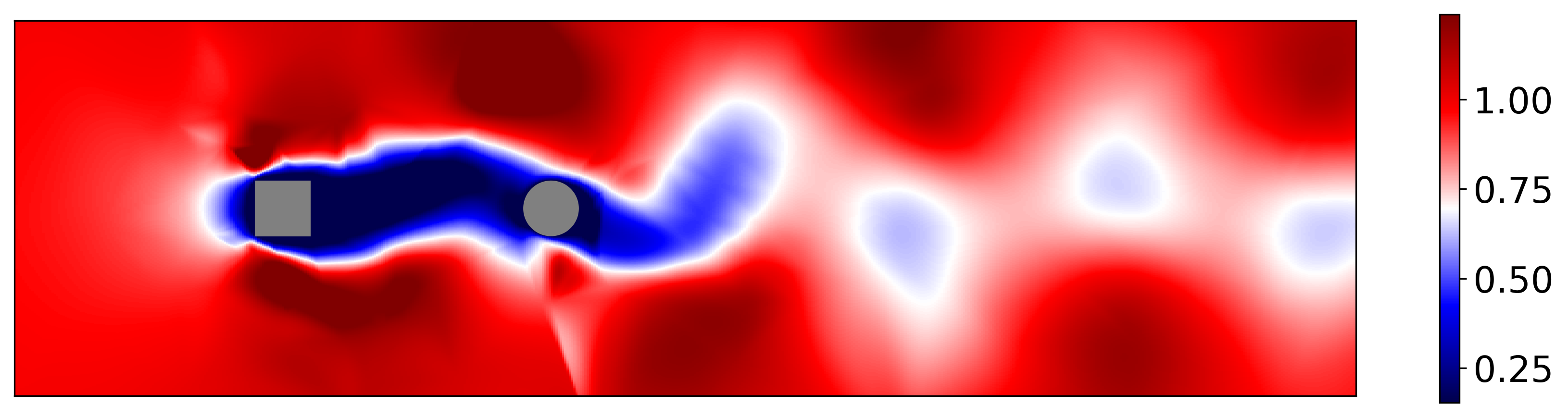}
        \caption{$(e)$ Actual $u_x$ at $t_0+20$}
    \end{subfigure}
        \begin{subfigure}[b]{0.46\textwidth}
        \centering
        \includegraphics[width=.98\hsize]{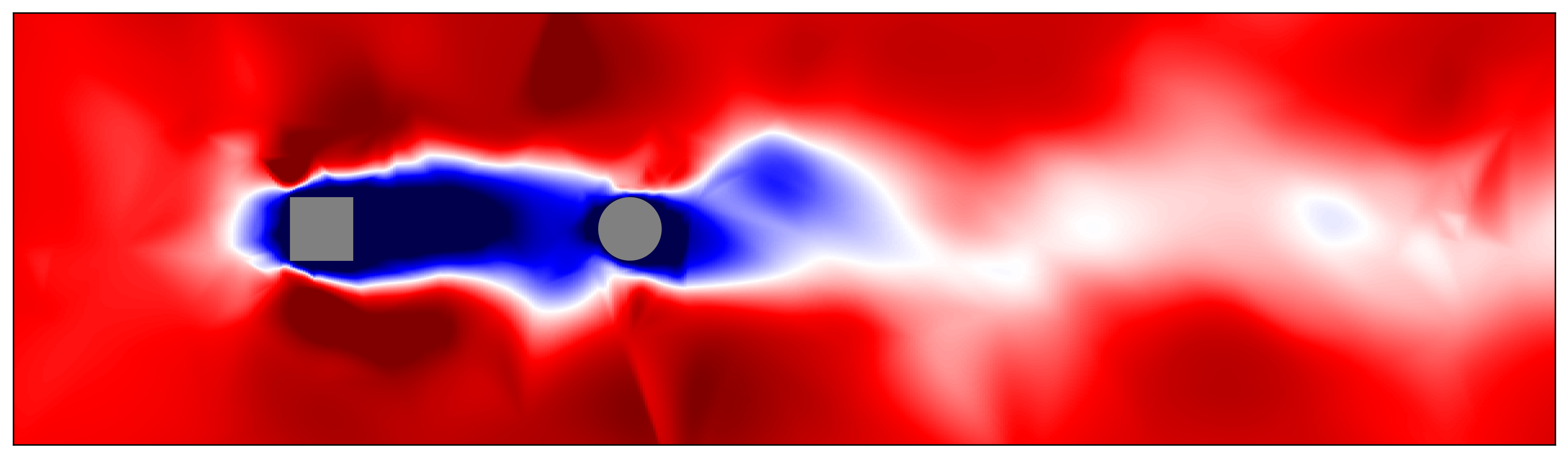}
        \caption{$(f)$ Reconstructed $u_x$ at $t_0+20$}
    \end{subfigure}
    \caption{The actual vs. reconstructed horizontal velocity $(u_x)$, field for the case of the wake induced free oscillation of the cylinder  at $U^*=15$ at times $t_0$ (a,b) , $t_0+10$ (c,d), and $t_0+20$ (e,f). }
    \label{fig:reconstructed_velocity_case1}
\end{figure*}

From the mapping, the value of $\tilde{\mathbb{Y}}_c$ is known, which allows extracting the estimated values of $\hat{\psi}_{i,e}$ and $\hat{\alpha}_{f,i,e}$ (combining the estimates of the real and imaginary components), and the estimated flow field eigenvalues $\hat{\lambda}_{f,i,e}$ are known because it is assumed that $\tilde{\lambda}_{f,i,e}=\tilde{\lambda}_{s,i}$. Using these values, it is possible to reconstruct an approximation of $Y$ using a similar reconstruction to that in \ref{eq:reconstruction}:
\begin{equation}
    \tilde{F}_{b+q}=\sum_{i = 1}^{a}{\hat{\psi}_{i,e} \tilde{\lambda}_{f,i,e}^q \hat{\alpha}_{f,i,e}},
\end{equation}
where $\tilde{F}$ is the predicted flow field.

\section{Results}

The modal characteristics of pressure fields obtained from the surface modes for the free oscillation problem are illustrated in Figure \ref{fig:mapped_modes_viv}. Overall, the reconstructed pressure field provides a high-fidelity representation of the essential characteristics of each mode, with only a moderate degree of error. Notably, the pressure oscillations observed downstream of the cylinder, indicative of vortex wake advection, are accurately captured in the reconstruction. This suggests that these modes may possess sufficient accuracy to properly reconstruct the entire flow field.

The reconstructed pressure and velocity fields are delineated in Figures \ref{fig:reconstructed_pressure_case1} and \ref{fig:reconstructed_velocity_case1}, respectively. The pressure field reconstruction generally exhibits a high level of accuracy, although the mapping fuses the discrete low-pressure zones induced by individual vortices into a single, heterogeneous low-pressure region. Importantly, the fidelity of the reconstruction appears to be temporally invariant within the evaluated 20-second time frame. This is despite this duration being ample for the advection of approximately three vortex pairs downstream. Notably, the predominant source of error emanates from the initial modal estimates rather than the time-projection of these modes. A consistent vortex wake smoothing effect is observed in the velocity field. The reconstruction demonstrates heightened accuracy in the regions proximate to the cylinder, which is congruent with the methodology that employs surface data from the cylinder for the reconstruction process.

\begin{figure*}[ht]
    \centering
    \captionsetup[subfigure]{labelformat=empty}
    \begin{subfigure}[b]{0.53\textwidth}
        \centering
        \includegraphics[width=.98\hsize]{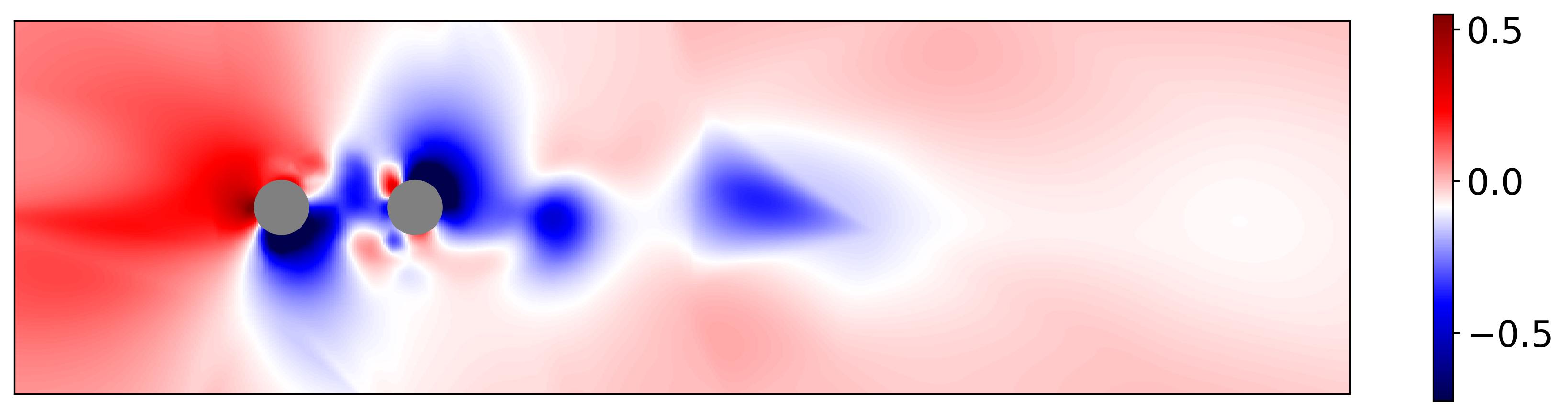}
        \caption{$(a)$ Actual pressure field at $t_0$}
    \end{subfigure}
        \begin{subfigure}[b]{0.46\textwidth}
        \centering
        \includegraphics[width=.98\hsize]{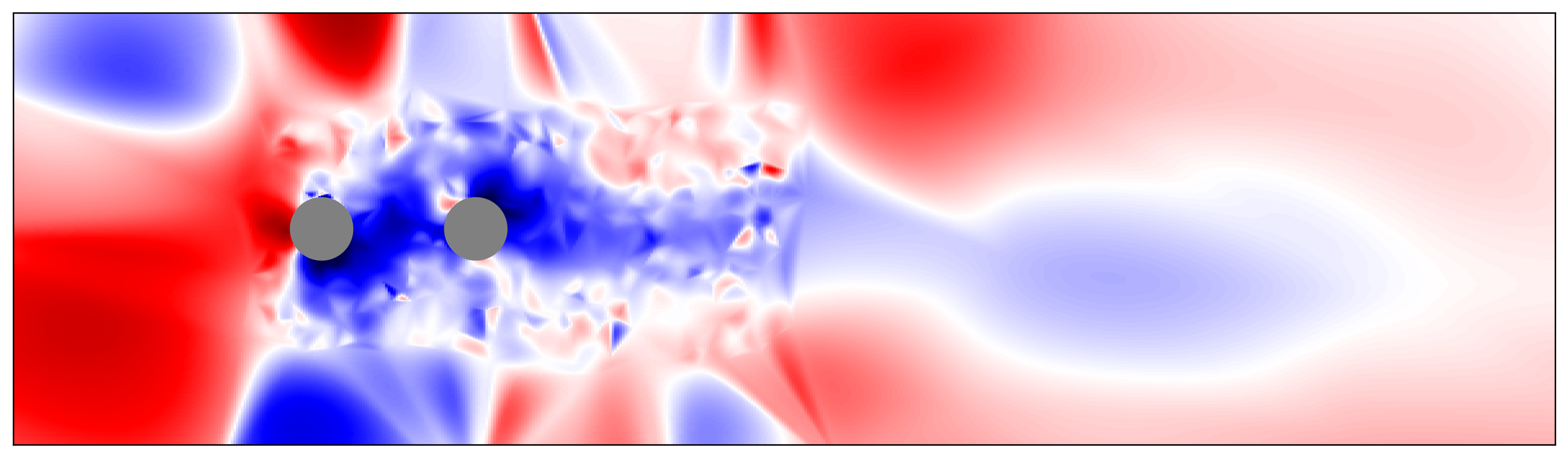}
        \caption{$(b)$ Reconstructed pressure field at $t_0$}
    \end{subfigure}
    \begin{subfigure}[b]{0.53\textwidth}
        \centering
        \includegraphics[width=.98\hsize]{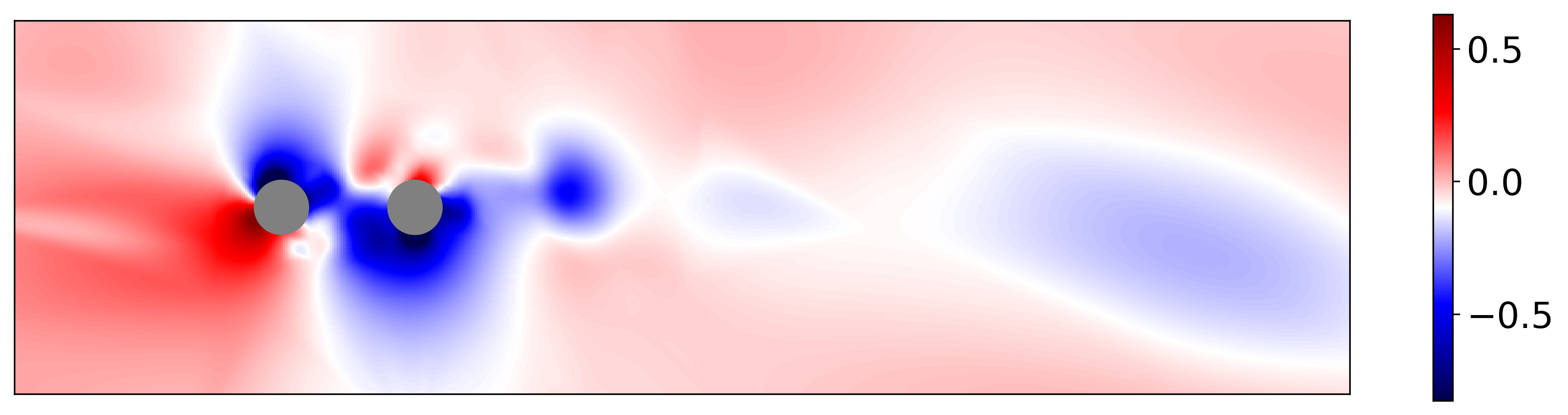}
        \caption{$(c)$ Actual pressure field at $t_0+10$}
    \end{subfigure}
        \begin{subfigure}[b]{0.46\textwidth}
        \centering
        \includegraphics[width=.98\hsize]{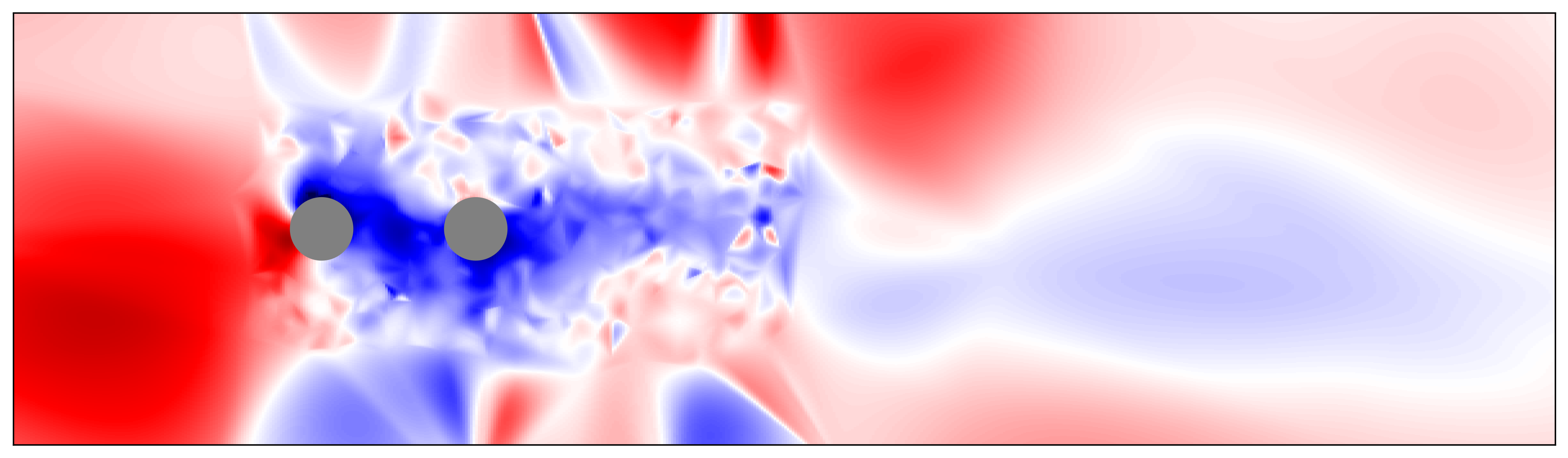}
        \caption{$(d)$ Reconstructed pressure field at $t_0+10$}
    \end{subfigure}
    \begin{subfigure}[b]{0.53\textwidth}
        \centering
        \includegraphics[width=.98\hsize]{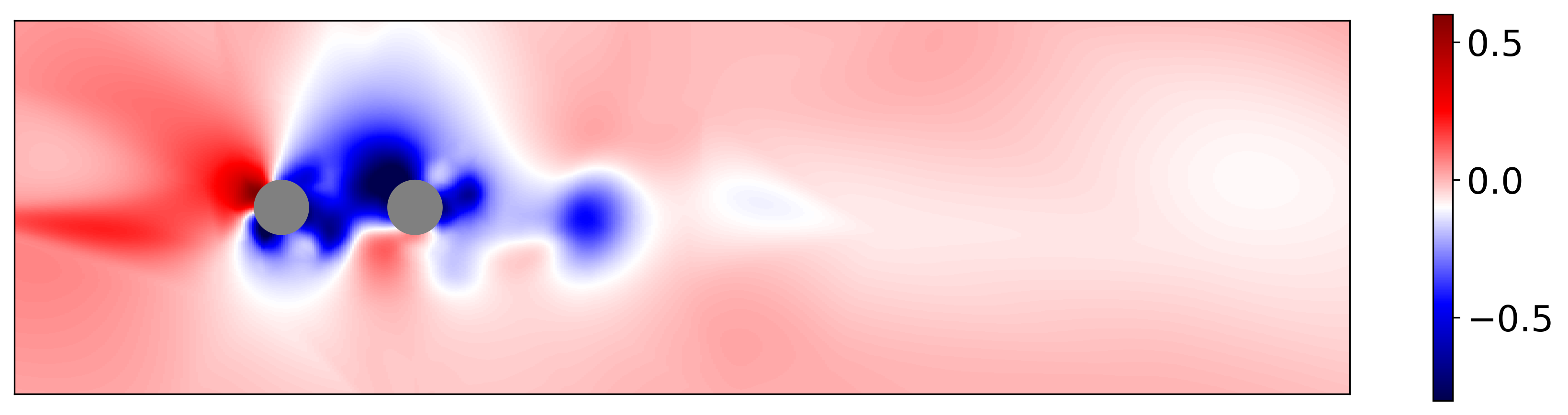}
        \caption{$(e)$ Actual pressure field at $t_0+10$}
    \end{subfigure}
        \begin{subfigure}[b]{0.46\textwidth}
        \centering
        \includegraphics[width=.98\hsize]{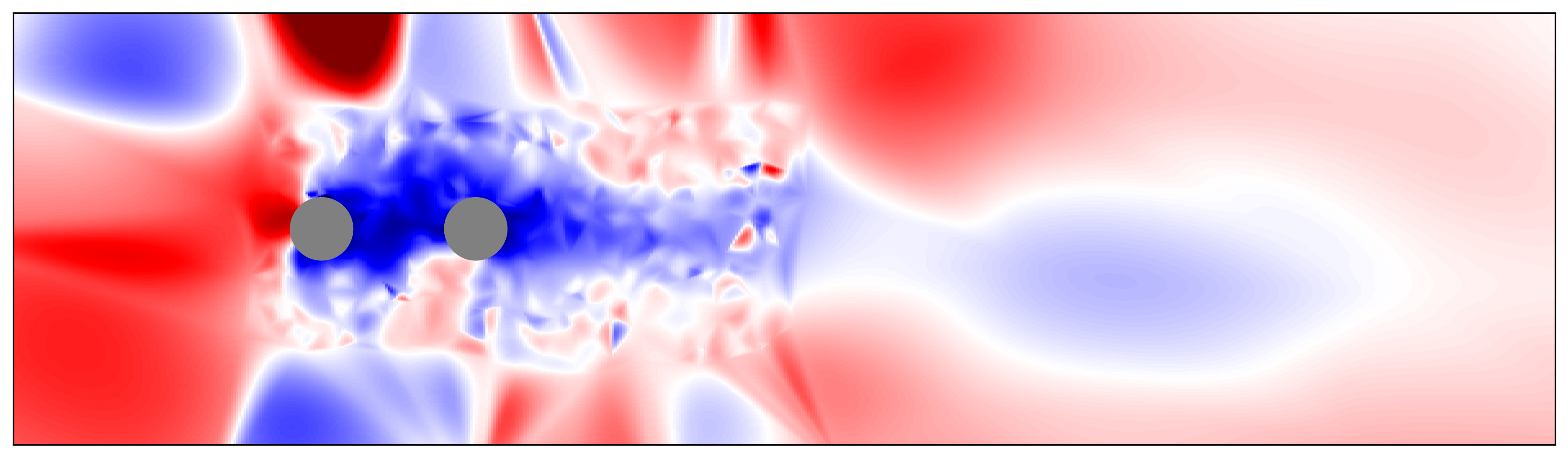}
        \caption{$(f)$ Reconstructed pressure field at $t_0+20$}
    \end{subfigure}
    \caption{The actual vs. reconstructed pressure field is depicted for the case of forced oscillations of two tandem cylinders, with an oscillation amplitude of \(A=0.5\) and a frequency of \(\omega = 1.04 \, \text{rad/s}\), at times \(t_0\) (a,b), \(t_0+10\) (c,d), and \(t_0+20\) (e,f).}
    \label{fig:reconstructed_pressure_case2}
\end{figure*}

\begin{figure*}[hbt!]
    \centering
    \captionsetup[subfigure]{labelformat=empty}
    \begin{subfigure}[b]{0.53\textwidth}
        \centering
        \includegraphics[width=.98\hsize]{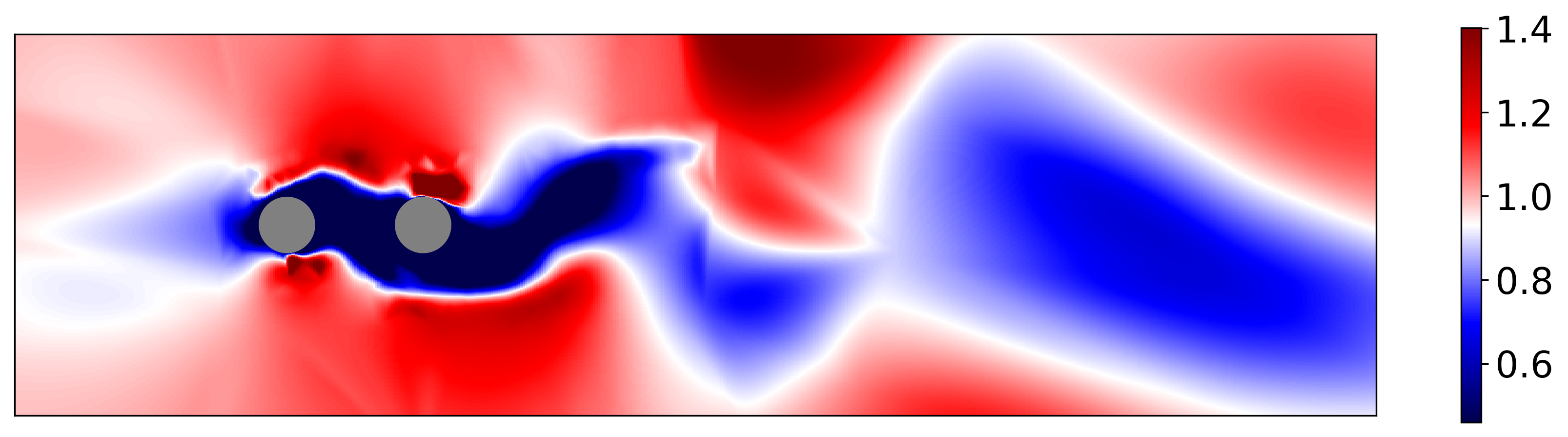}
        \caption{$(a)$ Actual $u_x$ at $t_0$}
    \end{subfigure}
        \begin{subfigure}[b]{0.46\textwidth}
        \centering
        \includegraphics[width=.98\hsize]{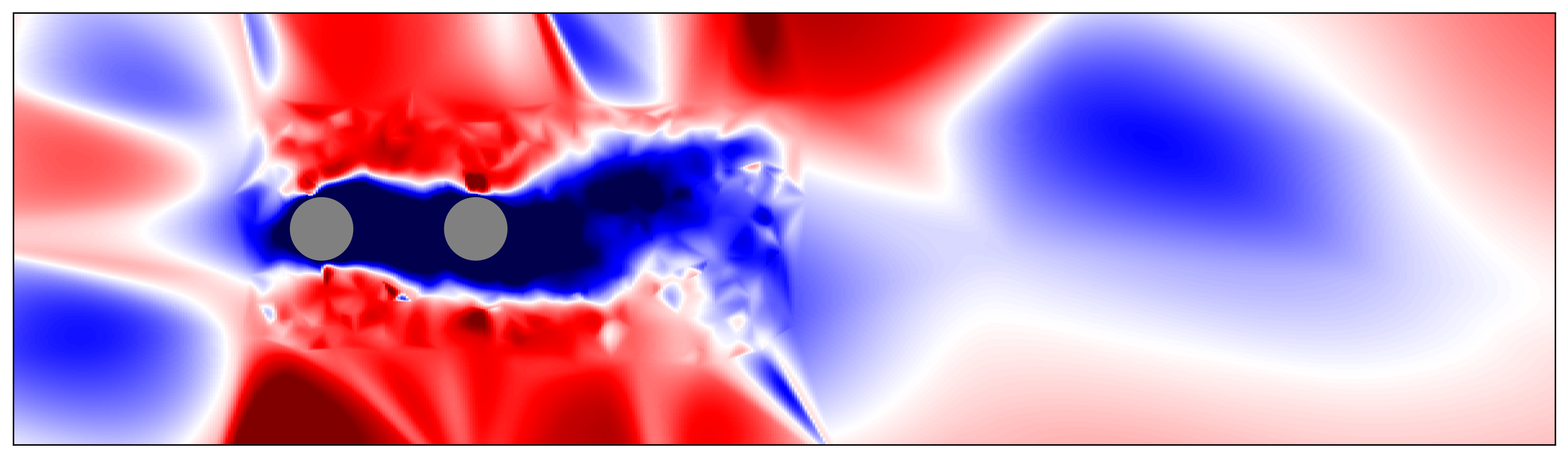}
        \caption{$(b)$ Reconstructed $u_x$ at $t_0$}
    \end{subfigure}
    \begin{subfigure}[b]{0.53\textwidth}
        \centering
        \includegraphics[width=.98\hsize]{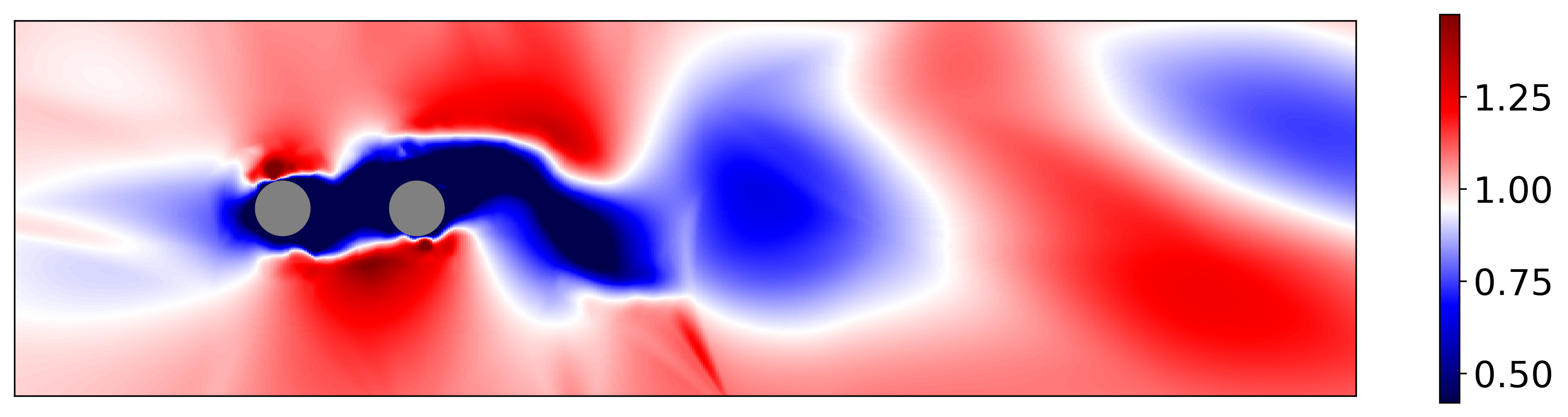}
        \caption{$(c)$ Actual $u_x$ at $t_0+10$}
    \end{subfigure}
        \begin{subfigure}[b]{0.46\textwidth}
        \centering
        \includegraphics[width=.98\hsize]{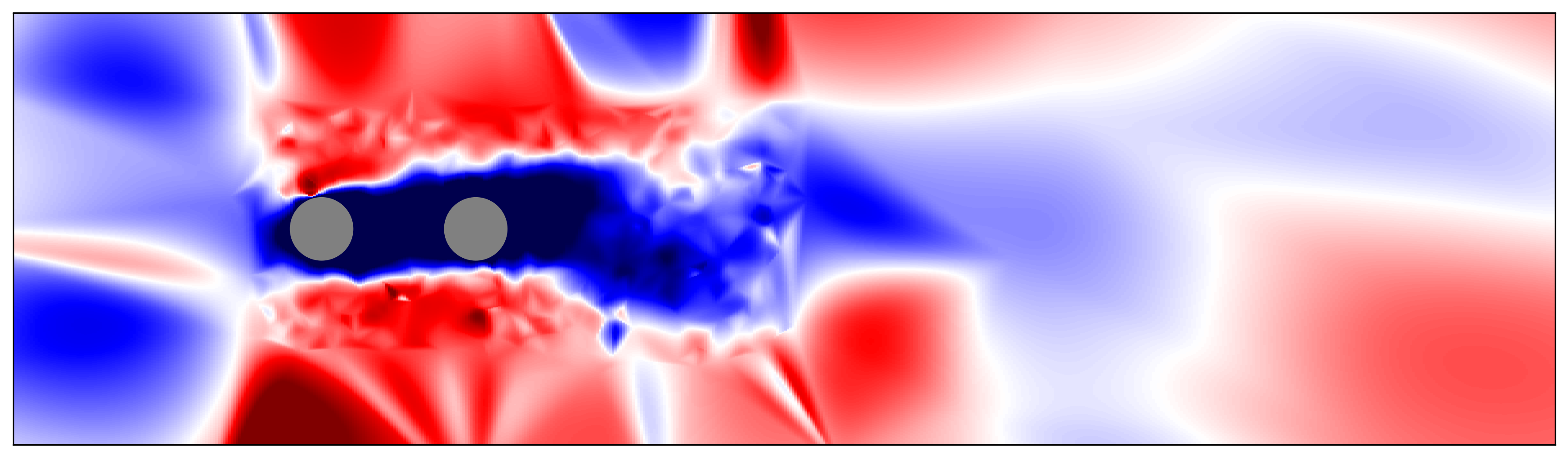}
        \caption{$(d)$ Reconstructed $u_x$ at $t_0+10$}
    \end{subfigure}
    \begin{subfigure}[b]{0.53\textwidth}
        \centering
        \includegraphics[width=.98\hsize]{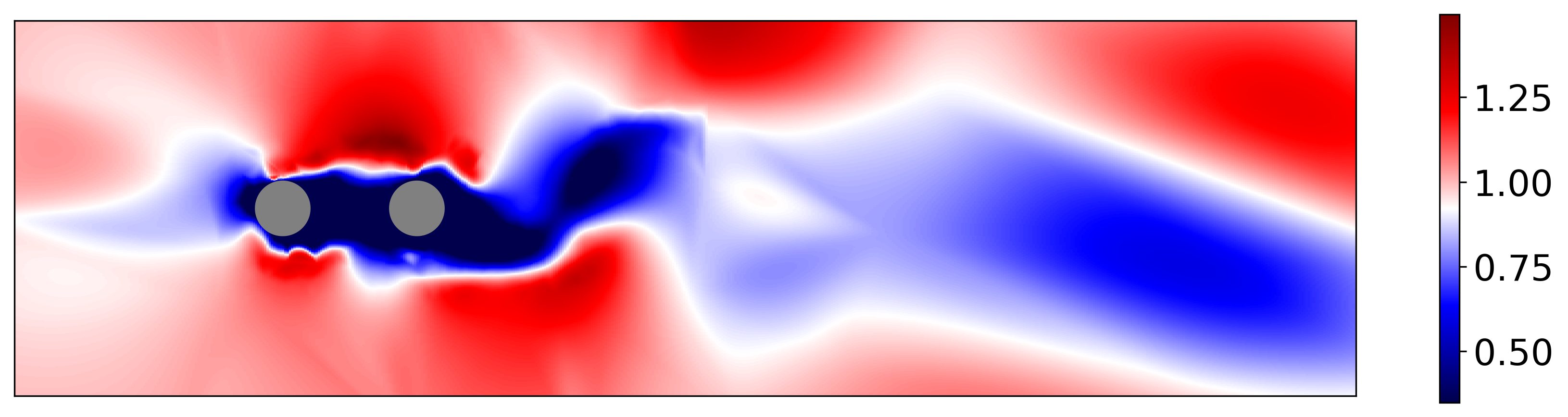}
        \caption{$(e)$ Actual $u_x$ at $t_0+20$}
    \end{subfigure}
        \begin{subfigure}[b]{0.46\textwidth}
        \centering
        \includegraphics[width=.98\hsize]{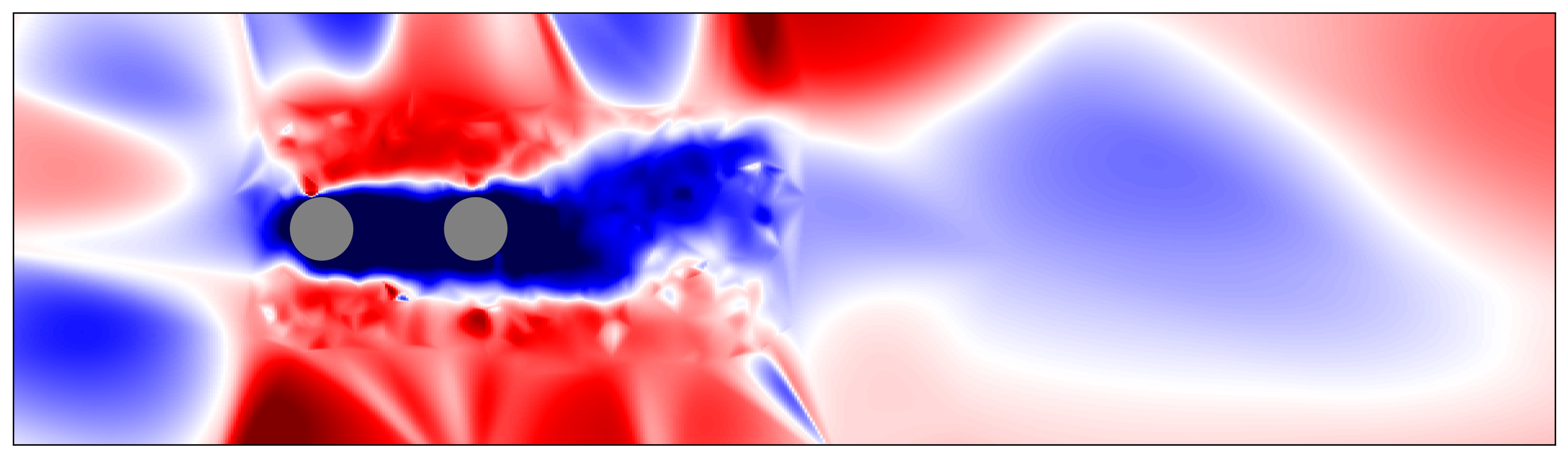}
        \caption{$(f)$ Reconstructed $u_x$ at $t_0+20$}
    \end{subfigure}
    \caption{The actual vs. reconstructed horizontal velocity ($u_x$) field is depicted for the case of forced oscillations of two tandem cylinders, with an oscillation amplitude of \(A=0.5\) and a frequency of \(\omega = 1.04 \, \text{rad/s}\), at times \(t_0\) (a,b), \(t_0+10\) (c,d), and \(t_0+20\) (e,f).}
    \label{fig:reconstructed_velocity_case2}
\end{figure*}

Reconstructing the forced oscillation case, where information about the flow is distributed across multiple modes, presents increased computational challenges. As depicted in Figures \ref{fig:reconstructed_pressure_case2} and \ref{fig:reconstructed_velocity_case2}, the reconstructed pressure and velocity fields exhibit certain idiosyncrasies. Notably, numerical noise is discernible in the rectangular region characterized by high mesh density adjacent to the bodies. The low-pressure oscillations localized on and between the cylinders are reconstructed with the highest fidelity, which is consistent with the data collection region serving as the basis for the flow field reconstruction. Comparative analysis suggests that the velocity field is reconstructed with greater accuracy than the pressure field. However, it should be highlighted that features located at a greater distance from the cylinders are generally subject to lower reconstruction accuracy than those in closer proximity.

\section{Conclusion}

In this study, we have established that Dynamic Mode Decomposition (DMD) can be effectively utilized to estimate the modes of a flow field using data  gathered exclusively from the surface of an immersed body. Moreover, by mapping both the magnitude and phase of these modes, we have successfully reconstructed the corresponding velocity and pressure fields. Our methodology has been demonstrated on two distinct fluid-body interaction scenarios: one involving free oscillations in the wake of a cylinder and the other encompassing forced oscillations. The approach has been demonstrated to be versatile, exhibiting applicability across both categories of fluid dynamics problems. These findings carry significant implications for underwater robotics, offering the potential to facilitate advanced features such as obstacle avoidance and optimal motion planning through an enhanced understanding of the surrounding fluid environment.

\begin{acknowledgments}
This work was supported by grant 13204704 from the Office of Naval Research and the National Science Foundation grant 2021612.
\end{acknowledgments}

\bibliography{Fish3}

\end{document}